\documentclass[sigconf]{acmart}

\usepackage[english]{babel}
\usepackage{blindtext}
\usepackage{color}
\definecolor{RED}{rgb}{1,0,0}
\definecolor{BLUE}{rgb}{0,0,1}
\definecolor{PURPLE}{rgb}{1,0,1}
\usepackage{enumitem}
\newcommand{\jiangkai}[1]{{#1}}

\newcommand{\wu}[1]{{#1}}

\usepackage{booktabs}
\usepackage{multirow}
\usepackage{makecell}
\usepackage{subcaption}
\usepackage{graphicx}
\usepackage{xurl}

\renewcommand\footnotetextcopyrightpermission[1]{} 
\setcopyright{none}

\acmDOI{}

\acmISBN{}

\acmPrice{}

\begin{document}
\title{Artic: AI-oriented Real-time Communication for MLLM Video Assistant}


\author{{Jiangkai Wu}, 
{Zhiyuan Ren}, 
{Junquan Zhong},
{Liming Liu},
{Xinggong Zhang}}
\affiliation{
  {Peking University}%
  \country{}%
}

\renewcommand{\shortauthors}{Jiangkai Wu et al.}

\begin{abstract}
AI Video Assistant emerges as a new paradigm for Real-time Communication (RTC), where one peer is a Multimodal Large Language Model (MLLM) deployed in the cloud. This makes interaction between humans and AI more intuitive, akin to chatting with a real person. However, a fundamental mismatch exists between current RTC frameworks and AI Video Assistants, stemming from the drastic shift in Quality of Experience (QoE) and more challenging networks. Measurements on our production prototype also confirm that current RTC fails, causing latency spikes and accuracy drops.

To address these challenges, we propose \textbf{\textit{Artic}}, an AI-oriented RTC framework for MLLM Video Assistants, exploring the shift from "humans watching video" to "AI understanding video." Specifically, Artic proposes: (1) Response Capability-aware Adaptive Bitrate, which utilizes MLLM accuracy saturation to proactively cap bitrate, reserving bandwidth headroom to absorb future fluctuations for latency reduction; (2) Zero-overhead Context-aware Streaming, which allocates limited bitrate to regions most important for the response, maintaining accuracy even under ultra-low bitrates; and (3) Degraded Video Understanding Benchmark, the first benchmark evaluating how RTC-induced video degradation affects MLLM accuracy. Prototype experiments using real-world uplink traces show that compared with existing methods, Artic significantly improves accuracy by 15.12\% and reduces latency by 135.31 ms. We will release the benchmark and codes at \href{https://github.com/pku-netvideo/DeViBench}{\textcolor{magenta}{https://github.com/pku-netvideo/DeViBench}}.
\end{abstract}

\maketitle

\section{Introduction}

\begin{figure}
\centering
\setlength{\abovecaptionskip}{0mm}
\includegraphics[width=0.8\linewidth]{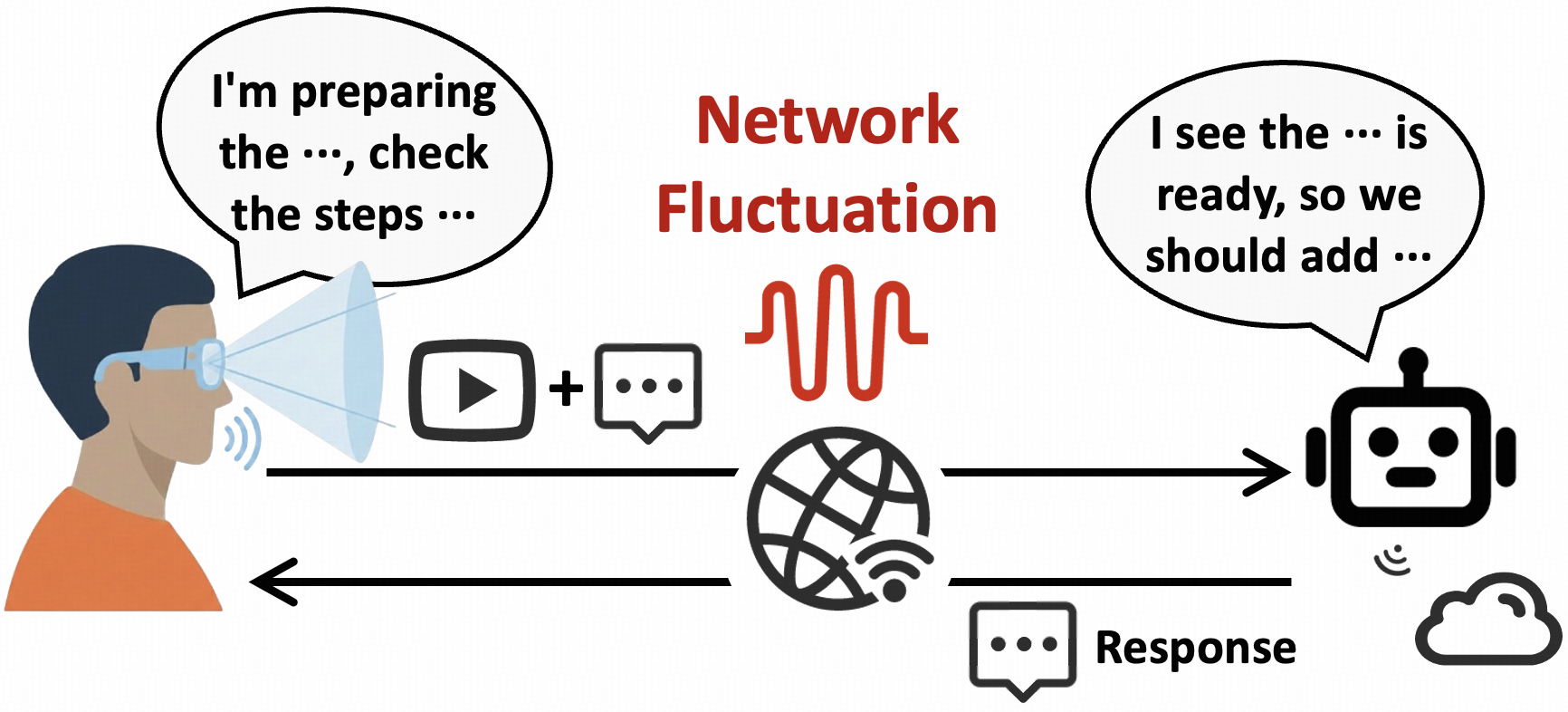}
\caption{AI Video Assistant is a new paradigm for real-time communication. The user sends video and audio to the AI for thinking. The AI feeds back audio.}
\label{fig:teaser}
\vspace{-1mm}
\end{figure}

AI Video Assistant~\cite{grok,chatgpt,gemini-live,copilot,doubao,yuanbao,tavus,soulfan,duolingo} is a new paradigm for Real-time Communication (RTC). Powered by Multimodal Large Language Models (MLLMs)~\cite{glm-realtime,xu2025qwen3omnitechnicalreport,tong2025interactiveomni,xu2025qwen2,team2024gemini,fu2025vita,zhang2024internlm}, these systems accept video and audio inputs directly, enabling intuitive interactions akin to chatting with a real human assistant~\cite{chen2024videollm}. Beyond basic question answering, modern assistants are evolving towards complex and agentic capabilities (\S\ref{sec:moti_background}), such as proactive response~\cite{project-astra,doubao,raven-0} and visual memory~\cite{chatgpt,gemini-live,glm-realtime,grok,replika,doubao}. However, MLLMs require high-performance computing resources (e.g., 8$\times$A100 GPUs~\cite{qiu2025modservemodalitystageawareresource}) to support real-time inference. Since AI Video Assistants are increasingly accessed via lightweight portable devices (e.g., smart glasses~\cite{omi-glass,ray-ban-meta-2,solo-airgo,inmo,galaxy-xr}) that cannot meet these demands, MLLMs are inevitably deployed on the server, relying on RTC to connect with the client. As shown in Figure~\ref{fig:teaser}, the client streams high-bitrate video and audio to the server while receiving low-bitrate audio and text responses.

However, a fundamental mismatch exists between current RTC frameworks~\cite{carlucci2016analysis} and AI Video Assistants, severely degrading the user Quality of Experience (QoE). First, the QoE objective has shifted from human perceptual quality (e.g., VMAF, stalling rate) to MLLM response accuracy and stricter latency constraints (\S\ref{sec:moti_diff}). Second, the usage scenario has evolved from sedentary applications (e.g., video conferencing) to "on-the-go" assistance (e.g., outdoor navigation~\cite{doubao,yang2026egolifeegocentriclifeassistant}), where mobility and unconstrained user behaviors induce drastic bandwidth fluctuations~\cite{narayanan2020lumos5g,feng2025vivisecting,hu2025comparative}. Third, the traffic pattern becomes uplink-dominant, making the system distinctively vulnerable to limited uplink capacity~\cite{feng2025vivisecting,ghoshal2025first,ghoshal:memu2022,guo2023power,khan2025mature,schippers2025donext} compared to traditional downlink-heavy (e.g., cloud gaming, remote desktop) or symmetric (e.g., video calls) services. Crucially, \textbf{measurements on our production prototype confirm that traditional human-oriented RTC fails in the wild, causing latency spikes and accuracy drops} (\S\ref{sec:moti_measure}) that shatter the user's illusion of interacting with a "real person." 


Is it possible to guarantee consistently low latency and robust high accuracy for AI Video Assistants \textbf{via RTC design}? \textit{Realizing this vision allows us to approach the "holy grail"~\cite{chen2024videollm} of AI research from a network systems perspective: making AI interact like a real human.} 

To achieve this, we identify three key challenges:

\textbf{First, Congestion Control (CC) becomes counterproductive upon accuracy saturation.} In traditional RTC, CC algorithms continuously increase video bitrate when bandwidth is available (Figure~\ref{fig:moti_latency}, left), as they are optimized for human eyes where higher bitrates consistently improve perceptual quality. In contrast, MLLMs exhibit a fundamentally different behavior: once video quality reaches a sufficient level (e.g., the 968 Kbps in Figure~\ref{fig:moti_accuracy}), further increasing bitrate yields negligible accuracy gains. Upon saturation, the traditional strategy of maximizing bitrate to fill the bandwidth is not only wasteful but dangerous; it occupies the bandwidth headroom, significantly increasing the risk of high latency when bandwidth fluctuates (e.g., frames 500--600 in Figure~\ref{fig:moti_latency}).

\textbf{Second, context-agnostic encoding is inefficient.} Low bandwidth compels RTC to encode video at low bitrates, which inevitably leads to video quality degradation and causes response errors (e.g., Figure~\ref{fig:moti_accuracy}(c)). We observe that to answer a specific question, MLLMs typically require high fidelity only in local regions. For instance, in Figure~\ref{fig:moti_accuracy}, when the request is "Please help me check the products I am restocking", only the products near the user's hands need to be legible; clarity in other areas has no impact on accuracy. However, existing context-agnostic encoding distributes bits indiscriminately, causing critical details to be blurred. While traditional Region of Interest (ROI) encoding can prioritize specific areas (e.g., faces in video conferencing), it relies on predefined targets. In contrast, an MLLM's focus is dictated by the current conversation context—which is highly dynamic and cannot be predetermined.

\textbf{Third, no existing benchmark evaluates how video degradation in RTC affects MLLM accuracy.} Traditional video streaming benchmarks focus on human perceptual quality (e.g., VMAF, stalling rate) rather than accuracy. Although some MLLM benchmarks evaluate response quality in streaming video scenarios (e.g., StreamingBench~\cite{lin2024streamingbench}), they aim primarily to test MLLM intelligence; thus, all input videos are ideally high-bitrate (e.g., 4000 Kbps). This renders existing QA samples too simple and high-level, requiring only coarse-grained video content to answer correctly. For instance, our tests on StreamingBench reveal that only 8\% of QA samples are degradation-sensitive (i.e., answered correctly at high bitrates but incorrectly at low bitrates). This fails to represent real-world scenarios, where detail-rich queries are highly susceptible to quality degradation. As shown in Figure~\ref{fig:moti_accuracy}, for the question "What is the text on the product?", even slight blurriness leads to incorrect responses.

To address these challenges, we propose \textbf{\textit{Artic}}, an \textbf{\textit{A}}I-oriented \textbf{\textit{r}}eal-\textbf{\textit{ti}}me \textbf{\textit{c}}ommunication system for MLLM Video Assistants, exploring the shift in network requirements from "humans watching video" to "AI understanding video." Artic fundamentally redesigns the RTC pipeline—spanning bitrate adaptation, video encoding, feedback control, benchmark construction, and QoE evaluation—to align network transmission with the MLLM's cognitive state. Overall, the system design of Artic entails three core contributions:

\begin{itemize}[left=0pt]
\item \textbf{Response Capability-aware Adaptive Bitrate.} Instead of blindly following CC, Artic incorporates the MLLM's response capability into bitrate adaptation. Specifically, Artic utilizes confidence feedback from the MLLM to determine if the current response capability is saturated. If so, Artic proactively caps the bitrate even when bandwidth is ample. This strategy reserves bandwidth headroom to absorb network fluctuations, minimizing latency spikes.

\item \textbf{Zero-overhead Context-aware Streaming.} Instead of context-agnostic encoding, Artic proposes allocating limited bitrate to regions most important to the current conversation context. Specifically, Artic instructs the MLLM to feed back regions essential for the response in real-time. Upon receiving the regions, the client adaptively adjusts encoding Quantization Parameters (QP), thereby ensuring the MLLM captures critical visual semantics even under ultra-low bandwidth. This MLLM-driven feedback avoids the overhead of client-side importance identification while offering more native and accurate context awareness.

\item \textbf{Degraded Video Understanding Benchmark.} Artic introduces the first benchmark evaluating how RTC-induced video degradation affects MLLM accuracy. Instead of manual annotation, Artic constructs an automated pipeline to generate "degradation-sensitive samples." Leveraging this pipeline, Artic produces 1,968 samples, totaling 88,680 seconds of video across 6*2 scene categories. This benchmark features a test and validation split, enabling not only QoE evaluation for AI Video Assistants but also hyperparameter tuning for the system and in-context learning for the MLLM. \textbf{We will open-source it.}

\end{itemize}
\vspace{0mm}
We prototyped and evaluated Artic with real-world mobile uplink traces. Compared to existing frameworks, Artic significantly improves accuracy by 15.12\% and reduces latency by 135.31 ms. We also validate Artic's generalizability and robustness. Furthermore, we demonstrate that Artic incurs minimal computation, bandwidth, and monetary overhead.

\section{Motivation}
\label{sec:moti}
\subsection{AI Video Assistant Background}
\label{sec:moti_background}

AI Video Assistants have rapidly emerged (e.g., Grok~\cite{grok}, Gemini Live~\cite{gemini-live}, ChatGPT~\cite{chatgpt}, Copilot~\cite{copilot} and Doubao~\cite{doubao}). Powered by Multimodal Large Language Models (MLLMs)~\cite{glm-realtime,xu2025qwen3omnitechnicalreport,tong2025interactiveomni,xu2025qwen2,team2024gemini,fu2025vita,zhang2024internlm}, these systems accept direct video and audio inputs rather than text prompts. The MLLM processes these inputs to generate responses, providing feedback to the user. This enables natural interactions comparable to human assistants. Beyond basic visual question answering, they are evolving towards more complex, agentic capabilities:

\noindent \textbf{Proactive Response.} Unlike early turn-based interactions responding only to explicit queries, modern assistants (e.g., Google Project Astra~\cite{project-astra}, Doubao~\cite{doubao}, raven-0~\cite{raven-0}) support full-duplex interactions. They proactively decide when to respond by analyzing environmental changes and user intent. For instance, if a user instructs "watch the game with me and shout 'goal' when a score occurs," the assistant continuously monitors the stream and responds timely upon the event.

\noindent \textbf{Visual Memory.} In contrast to early stateless interactions, current leading assistants (e.g., Gemini live~\cite{gemini-live}, ChatGPT~\cite{chatgpt}, Replika~\cite{replika}) incorporate memory capabilities. This ensures consistency across multi-round conversations. Furthermore, it enables the assistant to recall visual information from previous frames. For example, if a user asks, "I lost my phone, please tell me where I left it," the assistant can identify the phone in historical video frames and infer where it was lost.

\noindent \textbf{Lightweight \& Portable Clients.} AI Assistants are transitioning from stationary desktop platforms to mobile applications (e.g., Grok~\cite{grok}, Doubao~\cite{doubao}) and wearable devices (e.g., smart glasses~\cite{omi-glass,ray-ban-meta-2,solo-airgo,inmo,galaxy-xr}). This shift drives the deployment towards lightweight, portable terminals.

These capabilities necessitate RTC for AI Video Assistants. First, MLLMs require high-performance computing resources (e.g., 8$\times$A100 GPUs~\cite{qiu2025modservemodalitystageawareresource}) to support real-time inference. However, lightweight client devices cannot meet these computational and power demands. Consequently, MLLMs are inevitably deployed in the cloud, where the client transmits user video and audio for inference. Furthermore, to support proactive responsiveness, \textit{the system must maintain continuous RTC sessions lasting minutes to hours, rather than transmitting sparse images or short clips.} \textbf{Thus, the AI Video Assistant represents a novel scenario for RTC.}

\subsection{Differences between AI Video Assistant and traditional RTC}
\label{sec:moti_diff}

The unique characteristics of AI Video Assistants introduce novel network requirements for RTC, differing from human-centric communication. We analyze these shifts below.

\begin{table}[t]
\setlength{\abovecaptionskip}{2.mm}
\centering
\caption{Comparison of QoE Objectives}
\label{tab:qoe_comparison}
\renewcommand{\arraystretch}{1.6} 
\begin{tabular}{m{1.2cm}<{\centering} | m{3cm}<{\centering} | m{2.9cm}<{\centering}}
\hline
\textbf{Scenario} & \textbf{Traditional RTC} & \textbf{AI Video Assistant} \\
\hline
\multirow{5}{*}{\makecell{\textbf{QoE} \\ \textbf{factors}}} 
    & \makecell{\textbf{Fidelity} \\ \footnotesize (e.g., PSNR, SSIM, VMAF)} 
    & \multirow{4}{*}{\makecell{\textbf{Accuracy} \\ \footnotesize (MLLM responses)}} \\
\cline{2-2} 
    & \makecell{\textbf{Fluency} \\ \footnotesize (e.g., FPS, stalling rate)} & \\
\cline{2-2}
    & \makecell{\textbf{Consistency} \\ \footnotesize (e.g., quality fluctuations)} & \\
\cline{2-3} 
    & \makecell{\textbf{Latency} \\ \footnotesize (More budget allowed)} 
    & \makecell{\textbf{Latency} \\ \footnotesize (Less budget allowed)} \\
\hline
\end{tabular}
\end{table}

\noindent \textbf{QoE changes from human perceptual quality to MLLM response accuracy and latency.} In traditional RTC, QoE is centered on the human visual system. To satisfy human eyes, RTC systems optimize for Fidelity (using metrics like PSNR~\cite{zhang2021loki,zhang2020onrl}, SSIM~\cite{fouladi2018salsify,yan2020learning,cheng2024grace} or VMAF~\cite{vmaf} to ensure clarity), Fluency (maintaining high FPS~\cite{meng2022achieving,zhang2021loki,zhang2020onrl,rudow2023tambur} and minimizing stalling~\cite{cheng2024grace,zhang2021loki,zhang2020onrl,rudow2023tambur}), and Consistency (avoiding frequent quality fluctuations~\cite{zhang2020onrl}). \textit{However, in AI Video Assistants, the video consumer shifts from a human to an MLLM. Consequently, strict adherence to perception-related metrics is no longer necessary; instead, the primary optimization objective becomes the accuracy of the MLLM’s responses.} This elevates RTC’s tolerance to the semantic level. For example, as long as accuracy is maintained, visual degradations such as blurriness are acceptable (e.g., Figure~\ref{fig:moti_accuracy}(b)).

\textit{Although QoE requirements have relaxed in the above aspects, latency remains critical.} Since modern assistants~\cite{project-astra,doubao,raven-0} are designed to proactively respond the moment a target event occurs, any delay in frame delivery directly translates to a delay in the assistant's response. \textit{Furthermore, AI Assistants impose even stricter constraints on latency.} To ensure a fluent interactive experience, the response latency needs to remain below 300 ms~\cite{lai2022spacertc}. In traditional RTC, the human peer on the other side can respond instantly, meaning the latency budget is largely available for RTC. However, MLLMs generate responses in a time-consuming autoregressive manner. Even with audio-only input, computational latency reaches at least 232 ms~\cite{hurst2024gpt}. To keep the total response latency under 300 ms, the budget left for RTC is at most 68 ms, which is difficult to guarantee. Therefore, minimizing frame latency becomes the other primary objective alongside accuracy. We summarize these QoE factors in Table~\ref{tab:qoe_comparison}.

\begin{figure*}
\setlength{\abovecaptionskip}{1.mm}
    \centerline{\includegraphics[width=0.9\linewidth]{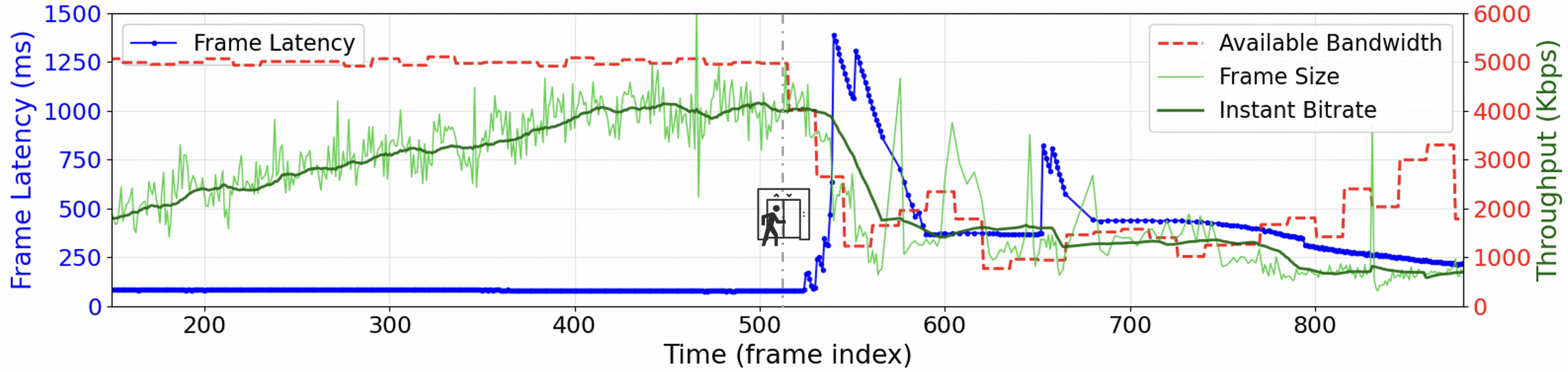}}
    \caption{Latency measurement study. When bandwidth is sufficient, the video bitrate continuously increases, as driven by the CC. However, influenced by mobility and user behaviors, bandwidth suffers sudden drops (e.g., frame 525). The lag in bitrate reduction causes network congestion, resulting in severe latency spikes (e.g., 1,389 ms).}
    \label{fig:moti_latency}
\end{figure*}

\noindent \textbf{High mobility and user behavior exacerbate bandwidth fluctuations.} The portability of AI Video Assistants (e.g., smart glasses~\cite{omi-glass,ray-ban-meta-2,solo-airgo,inmo,galaxy-xr}) necessitates "on-the-go" usage, introducing significant stability challenges to RTC. Physical mobility inherently degrades link quality. Whether on Wi-Fi or 5G, continuous movement triggers frequent signal attenuation and handovers~\cite{narayanan2020lumos5g,feng2025vivisecting,hu2025comparative}, resulting in drastic bandwidth fluctuations exceeding static connections.

\textit{More critically, user behavior becomes unconstrained.} For example, in traditional video conferencing, users are often cooperative; they proactively seek locations with stable connectivity and remain stationary to ensure network quality. In contrast, AI Video Assistants are used "in the wild." Users expect seamless interaction while navigating complex environments—such as entering elevators or subways—subjecting the RTC session to drastic bandwidth drops and outages that are rarely encountered in static settings.

\noindent \textbf{Uplink is more pressing than Downlink.} 
Traditional RTC scenarios typically exhibit downlink-heavy or symmetric traffic patterns. In 1v1 video calls, traffic is symmetric as each peer acts as both a video sender and receiver. In video conferencing, most participants (except the active speaker) face higher downlink pressure, receiving multiple streams while sending only one. Similarly, in interactive video like cloud gaming or remote desktop, clients transmit low-bitrate control signals while receiving high-quality video, making the traffic heavily downlink-dominant. These patterns align well with current network infrastructures (e.g., 5G), where downlink bandwidth is designed to be significantly higher than uplink~\cite{feng2025vivisecting,ghoshal2025first,ghoshal:memu2022,guo2023power,khan2025mature,schippers2025donext}, such as 4.8x--6.5x~\cite{feng2025vivisecting}. 

\textit{In contrast, AI Video Assistants rely on unidirectional uplink transmission.} The client acts exclusively as the video sender, continuously streaming high-bitrate video to the MLLM, while the MLLM returns feedback only via low-bitrate audio or text. This traffic pattern runs contrary to the asymmetric design of modern networks. Consequently, the bottleneck shifts entirely to the uplink, imposing greater challenges on QoE compared to traditional RTC scenarios.

\vspace{-0mm}
\subsection{Measurement study: Real-world Network Impair Assistant QoE}
\label{sec:moti_measure}

To empirically demonstrate how real-world network limitations impair the AI Assistant experience, we conducted a measurement study in this section. Next, we will first describe the measurement setup and then present the analysis.

To capture the network challenges in \S\ref{sec:moti_diff}, we collected real-world network traces that reflect realistic traffic patterns. We established a testbed using an iPhone 17 on a commercial 5G network, generating uplink UDP traffic via iperf~\cite{iperf} to a production linux server. We logged the receiving throughput to estimate the available bandwidth. To rigorously isolate the impact of mobility from other factors, we employed two specific strategies. First, we conducted tests at 2:45 AM to minimize cross-traffic contention. Second, we capped the sending rate at 5 Mbps—a rate the link could easily sustain when static—to eliminate internet-side bottlenecks. This ensures that any observed fluctuations are attributable to physical environment changes rather than external network congestion. As for user behavior, we selected taking an elevator, a ubiquitous scenario in daily life.

\begin{figure*}
\setlength{\abovecaptionskip}{1.mm}
    \centerline{\includegraphics[width=\linewidth]{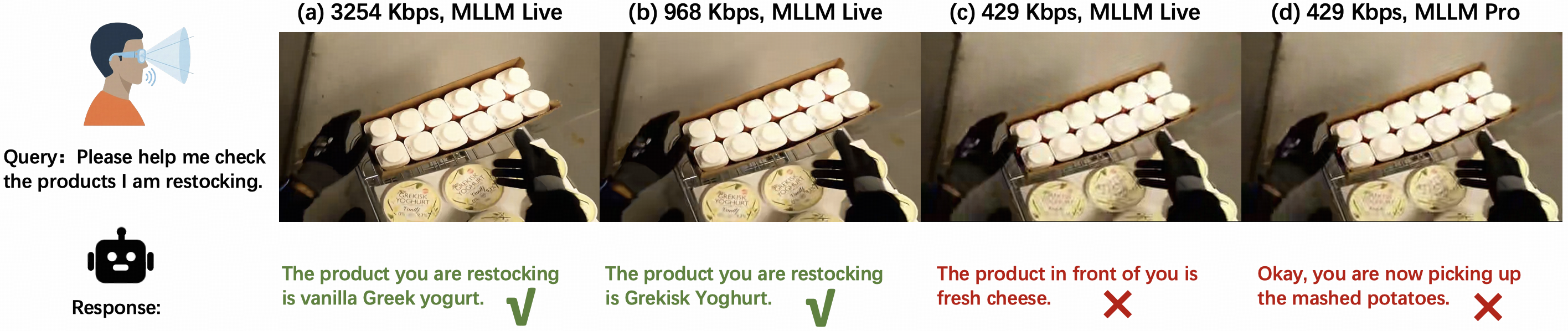}}
    \caption{Accuracy measurement study. (a)(b) Accuracy exhibits saturation. The MLLM remains consistently accurate as the bitrate increases. (c) Low bitrates cause MLLM errors. (d) Errors persist even with the more powerful Pro model, \textit{demonstrating that errors stem from video degradation rather than model capability.}}
    \vspace{-0mm}
    \label{fig:moti_accuracy}
\end{figure*}

We implemented an AI Video Assistant prototype following OpenAI's AI video call standards~\cite{openai_rtc}. The frontend is a web-based AI assistant acting as the RTC sender, capturing user audio and video. The backend runs on a production linux server as the RTC receiver, invoking the MLLM for responses. Specifically, we utilized the commercial Agora SDK~\cite{agora_pricing} for RTC and Google Gemini Live API~\cite{gemini-live} for the MLLM. To control network conditions, we routed the sender's traffic through a custom pipe that dynamically limits bandwidth to replay the collected traces.

We visualize the bandwidth trace, video bitrate, frame size, and frame latency in Figure~\ref{fig:moti_latency}, and accuracy examples in Figure~\ref{fig:moti_accuracy}. Based on these results, we make the following observations:

\noindent \textbf{Challenging network conditions are the norm.} As shown in the Figure~\ref{fig:moti_latency}, when the user is static, the available bandwidth consistently saturates the 5 Mbps limit, confirming a stable, sufficient link under ideal conditions. However, at frame 525, when the user enters the elevator, the available bandwidth rapidly collapses to 1.23 Mbps. \textit{This demonstrates that for "on-the-go" AI Video Assistants, network quality is dictated by routine user behaviors and physical environments}, shifting from excellent to poor within 1.5 seconds.

\noindent \textbf{High bandwidth fluctuations lead to latency spikes.} In RTC systems, bitrate is typically adapted by CC algorithms (e.g., GCC~\cite{carlucci2016analysis}, BBR~\cite{cardwell2017bbr}). When latency or packet loss does not increase, the CC algorithm ramps up the bitrate to probe for available bandwidth. \textit{However, when the aforementioned sudden bandwidth drop occurs, CC algorithm exhibits an adaptation lag (approximately 1.17 seconds)}. Consequently, the bitrate exceeding the bandwidth leads to severe packet accumulation, resulting in a latency as high as 1,389 ms.

\noindent \textbf{Low bandwidth degrades accuracy.} In response to high latency and packet loss, the CC algorithm eventually slashes the bitrate. However, as illustrated in Figure~\ref{fig:moti_accuracy}, this low bitrate results in severe blurring, which directly causes incorrect responses from the MLLM. To verify that these errors stem from video degradation rather than model's limited intelligence, we replaced the Gemini Live model with the Gemini 3 Pro~\cite{gemini_pro} model processing the same degraded frames. The Pro model persisted in making the same errors, confirming that \textit{the accuracy drop is rooted in the RTC-induced visual artifacts rather than the model's capability}.

\subsection{Key Insights and Potential Gains}
\label{sec:moti_insight}

Driven by the measurements above, our goal is to optimize AI Assistant QoE from an RTC perspective, particularly under challenging network conditions. This section presents three key insights derived from the differences between human and MLLM perception, followed by potential gains:

\noindent \textbf{The saturation of MLLM accuracy allows for latency reduction during bandwidth fluctuations.} Traditional RTC algorithms aggressively increase video bitrate when bandwidth is sufficient, optimized for human eyes where higher bitrates consistently improve perceptual quality (e.g., enabling 8K@120FPS). However, as observed in our measurements (Figure~\ref{fig:moti_accuracy}),  MLLMs exhibit a fundamentally different behavior: Once the video quality reaches a sufficient level (e.g., 968 Kbps), the model's answers always remain correct. In this saturation regime, further increasing bitrate yields no gains in accuracy. Instead, it occupies the bandwidth headroom, significantly increasing the risk of high latency when bandwidth fluctuates.

Instead of blindly increasing bitrate, we incorporate the model's response capability into the bitrate decision loop. If the MLLM can respond correctly at the current bitrate, we stop increasing bitrate even with ample bandwidth (e.g., frames 0--500 in Figure~\ref{fig:moti_latency}). This strategy deliberately maintains a bitrate "maximum margin," reserving headroom to absorb future bandwidth fluctuations (as seen at frame 525). 

\noindent \textbf{Context-aware coding improves accuracy under low bandwidth.} In video coding, a trade-off between bitrate and distortion always exists. Low bitrate causes degradation like blurriness and blocking artifacts. In traditional RTC, a viable solution is to allocate more bitrate to the Region of Interest (ROI) for human eyes (e.g., faces in video conferencing), sacrificing other areas to preserve the key regions. 

We observe that this concept is equally critical for AI Video Assistants, but the importance of different regions depends on the current conversation context rather than human gaze. For example, in Figure~\ref{fig:moti_accuracy}, given the query "Please help me check the
products I am restocking", the MLLM only needs to focus on the regions of products near the user’s hands, while the rest is irrelevant. Thus, rather than coding in a context-agnostic manner, the video should be context-aware. More bitrate should be allocated to response-important regions, and less to irrelevant ones.

\noindent \textbf{The first benchmark evaluates how video streaming quality affects MLLM accuracy.} According to \S\ref{sec:moti_diff}, QoE metrics in AI Video Assistant change from perception to accuracy. Existing video streaming benchmarks are inapplicable, as they focus on perceptual quality and do not involve response accuracy. In the MLLM field, there are some benchmarks targeting Streaming Video Understanding tasks~\cite{wang2025omnimmi,yangsvbench}, such as StreamingBench~\cite{lin2024streamingbench}. In these benchmarks, each video includes several Question-Answer (QA) samples for evaluating the response accuracy. However, these benchmarks aim to test the MLLM's intelligence, so all the input videos are ideally high-bitrate (e.g., 4000 Kbps). 

To evaluate how video streaming quality affects accuracy, we transcode videos from StreamingBench to 200 Kbps. 
We test on these low-bitrate videos with the original QA samples. The results show that only \textbf{8\%} of QA samples are answered incorrectly at low bitrate and correctly at high bitrate. This is because the QA samples in StreamingBench are too simple and high-level, requiring only coarse-grained video content to answer correctly. 
However, in real-world scenarios, there are often many detail-rich questions that are very sensitive to video quality. For example, in Figure~\ref{fig:moti_accuracy}, when the question is "What is the text on the product?", even slight blurriness will prevent correct answers. So it is necessary to establish a more challenging benchmark to reflect the real-world impact of video degradation on MLLM accuracy.

\vspace{0mm}
\section{Overview}
\label{sec:overview}
Based on the above insights, we propose \textbf{\textit{Artic}}, an \textbf{\textit{A}}I-oriented \textbf{\textit{r}}eal-\textbf{\textit{ti}}me \textbf{\textit{c}}ommunication system for Video Assistants. As shown in Figure~\ref{fig:overview}, Artic comprises three core modules: Response Capability-aware Adaptive Bitrate (\S\ref{sec:abr}), Zero-overhead Context-aware Streaming (\S\ref{sec:roi}), and the Degraded Video Understanding Benchmark (\S\ref{sec:bench}). Its workflow is as follows:

\begin{figure}
\setlength{\abovecaptionskip}{3.mm}
    \centerline{\includegraphics[width=\linewidth]{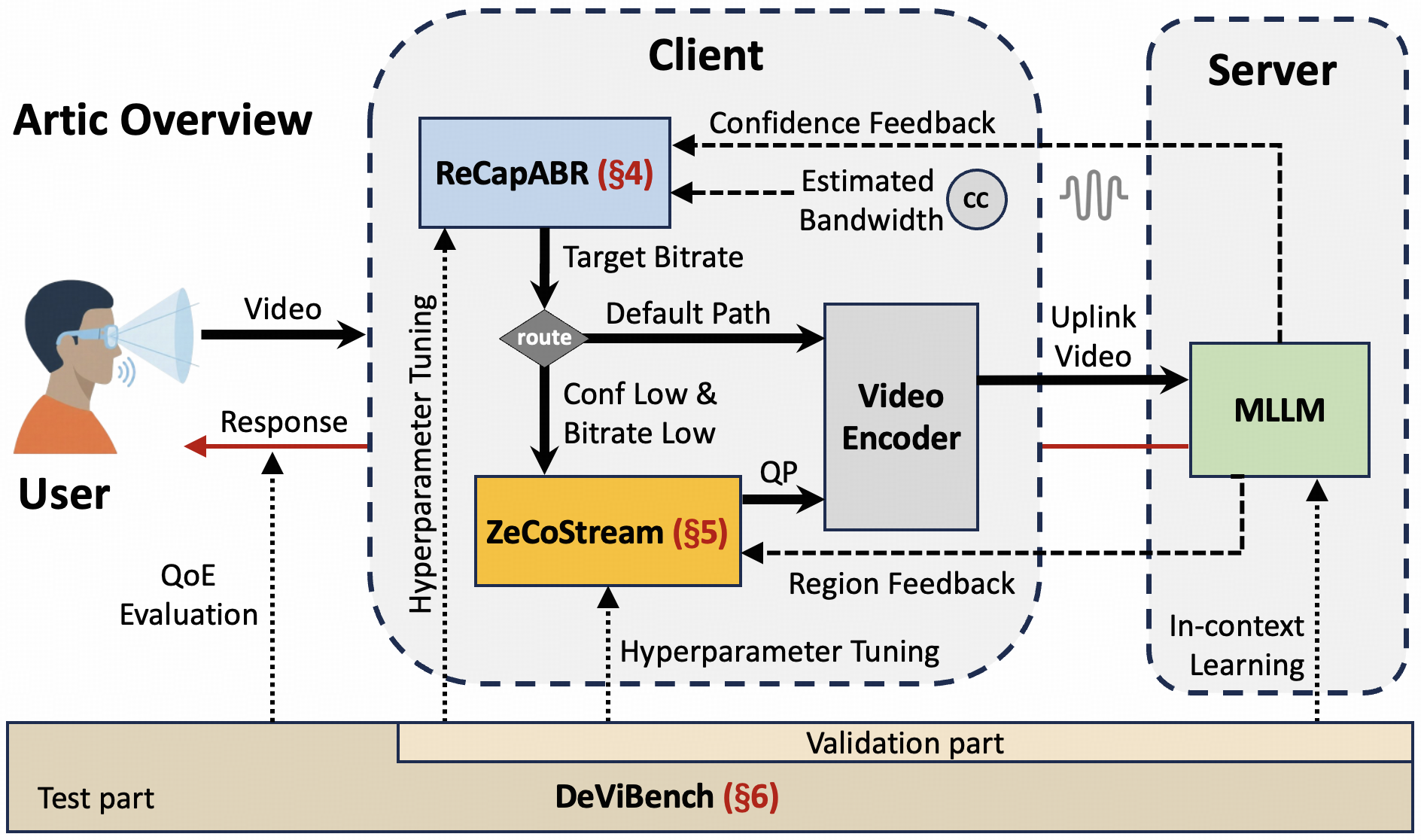}}
    \caption{Artic overview}
    \vspace{-3mm}
    \label{fig:overview}
\end{figure}

\noindent \textbf{\textit{ReCapABR}} (\textbf{\textit{Re}}sponse \textbf{\textit{Cap}}ability-aware \textbf{\textit{A}}daptive \textbf{\textit{B}}it\textbf{\textit{r}}ate) determines the video bitrate by bridging network CC with the MLLM's cognitive state. It utilizes the confidence score fed back from the MLLM to assess whether the model's current response capability is saturated. Unlike traditional ABR, ReCapABR introduces a retention mechanism: if the CC suggests increasing the bitrate (due to available bandwidth) but the MLLM indicates it is already performing well at the current quality, ReCapABR overrides the CC to maintain the current bitrate. This strategy reserves bandwidth headroom to absorb future fluctuations. In other cases—such as when the MLLM struggles to answer accurately or bandwidth declines—ReCapABR follows the CC's target bitrate to maximize video quality while avoiding congestion.

\noindent \textbf{\textit{ZeCoStream}} (\textbf{\textit{Ze}}ro-overhead \textbf{\textit{Co}}ntext-aware \textbf{\textit{Stream}}ing) optimizes spatial bitrate allocation. Under normal network conditions, the bitrate determined by ReCapABR is fed into a standard video encoder. However, when the bitrate drops to a critical level where the MLLM struggles to answer accurately, and the bandwidth does not permit a higher bitrate, ZeCoStream is triggered. Based on the feedback from the MLLM, ZeCoStream becomes aware of "response-important" regions without introducing extra overhead. Then, it directs the encoder to perform QP-adaptive encoding, such that the limited bitrate is allocated almost exclusively to these regions. 
This ensures that the MLLM captures critical visual semantics even under ultra-low bandwidth constraints. ZeCoStream is disabled by default, as the degradation of response-irrelevant regions leads to incomplete memory, potentially compromising future conversational contexts. Thus, it is triggered only in the necessary cases described above.

\noindent \textbf{\textit{DeViBench}} (\textbf{\textit{De}}graded \textbf{\textit{Vi}}deo Understanding \textbf{\textit{Bench}}mark) serves as the evaluation foundation. We construct DeViBench to quantitatively evaluate how RTC-induced degradation impacts MLLM accuracy. It comprises a test set and a validation set. The test set is used to evaluate the QoE of the Artic system, while the validation set facilitates the hyperparameter tuning of ReCapABR and ZeCoStream, as well as the in-context learning of the MLLM.
\section{Response Capability-aware Adaptive Bitrate}
\label{sec:abr}

In this section, we propose \textbf{ReCapABR}. We first analyze factors affecting MLLM accuracy under different bitrates. Second, we explore how to achieve response capability awareness. Third, we present the confidence-based bitrate control.

\subsection{What factors affect MLLM accuracy under different bitrates?}
\label{sec:abr_factor}

Unlike human perception, degrading with bitrate, MLLM understanding exhibits "saturation". We identify three key factors determining MLLM accuracy at low bitrates:

\noindent \textbf{Prior Knowledge vs. Out-of-Distribution (OOD) Content.} MLLMs are pre-trained on massive datasets, embedding vast internal knowledge. When visual inputs align with this prior knowledge, the model can infer the correct result even with degraded inputs. For example, when watching a World Cup match, if the MLLM detects a specific jersey color and a number, it can infer the player's identity (e.g., "Messi") based on its internal knowledge, even if the face is completely unrecognizable due to low bitrate blur. Conversely, for OOD content, the model cannot rely on priors and strictly requires high visual fidelity to extract features.

\noindent \textbf{Memory of Seen vs. Unseen Content.} As discussed in §2.1, current MLLMs maintain visual memory for a period. If content has been captured clearly in the past, the model can "recall" its attributes from memory, reducing the dependency on the current frame's quality. For example, if a user holds up a medicine bottle and the camera captures a clear frame of the label, the MLLM stores this information. Minutes later, even if the video becomes blurry due to a network drop, the MLLM can still answer questions about the dosage by retrieving from its memory, whereas newly appeared content would require immediate high quality to be recognized.

\noindent \textbf{Low vs. High Information Density.} The density of semantic information within the frame dictates the impact of compression artifacts. Scenes with sparse semantic information are resilient, while dense scenes are fragile. For example, a scene showing an outdoor lawn and sky remains semantically unchanged even when blurry, as the information loss does not alter the fundamental understanding of the scene. In contrast, a text-heavy document or a screen filled with code relies on high-frequency details. These details are the first to be smoothed out by video codecs at low bitrates, leading directly to incorrect recognition or hallucinations.

\subsection{Response capability awareness}
\label{sec:abr_aware}

To achieve response capability awareness, the most straightforward approach is to construct a mathematical model mapping bitrate and the above influencing factors to MLLM accuracy, similar to traditional QoE modeling~\cite{guan2019pano,wu2023zgaming}. However, explicitly establishing such a mapping model poses significant challenges in the context of AI Video Assistants:

\noindent \textbf{Generalization.} MLLMs exhibit extreme heterogeneity. Different model families (e.g., ChatGPT vs. Gemini) and parameter sizes (e.g., 7B vs. 70B) result in vastly different sensitivities to video degradation. A regression model trained on one MLLM is unlikely to transfer effectively to another. In addition, frequent commercial MLLM updates rapidly invalidate offline-trained predictors.

\noindent \textbf{Complexity.} As analyzed above, the relationship between visual quality and response accuracy is highly non-linear and content-dependent (e.g., relying on prior knowledge or memory), making lightweight fitting functions ineffective.

To address these challenges, we bypass external modeling and leverage the MLLM's intrinsic capabilities. Since the cloud-side MLLM already possesses strong video understanding and reasoning abilities, a feasible approach is to let it directly evaluate and feedback its response capability under current bitrates. Specifically, we employ in-context learning to establish a real-time feedback loop. By designing a specific system prompt, we instruct the MLLM to assess the information sufficiency of the currently received video. The MLLM then outputs a Response Confidence Score $C_t \in [0, 1]$ for the current timestamp $t$, which is fed back to the client.

\subsection{Confidence-based bitrate control}
\label{sec:control}
Based on the real-time confidence feedback $C_t$, we design a bitrate control algorithm to maintain the bitrate at a level just sufficient for the MLLM to respond, thereby reserving bandwidth headroom to absorb bandwidth fluctuations.

We first quantify the difference between the current performance and the target. Let $\tau$ be the confidence threshold (we set $\tau=0.8$ by the validation set \S\ref{sec:bench_summary}), which represents the confidence level required for the MLLM to respond correctly. We define the normalized confidence gap as $\delta_t = (\tau - C_t) / \tau$. A positive $\delta_t$ indicates that the MLLM struggles to recognize video details, requiring higher quality, while a negative $\delta_t$ suggests that the MLLM can already respond well with the current bitrate. To map this gap to the control logic, we compute a weight $w_t$ using a sensitivity parameter $\gamma$:
\begin{alignat}{2}
w_t=\delta_t \cdot\left|\delta_t\right|^{\gamma-1}
\label{equ:w_t}
\end{alignat}

We set $\gamma = 2$ by validation set \S\ref{sec:bench_summary}. This quadratic scaling allows the algorithm to react aggressively when the confidence is significantly low to restore accuracy quickly, while performing fine-grained adjustments as $C_t$ approaches $\tau$.

Finally, we employ $w_t$ to update the bitrate. Let $\hat{B}t$ be the estimated bandwidth from the CC algorithms. The next-step bitrate $R_{t+1}$ is calculated as:
\begin{alignat}{2}
R_{t+1}=\min \left(\hat{B}_t, R_t+w_t \cdot\left(\hat{B}_t-R_t\right)\right)
\label{equ:abr}
\end{alignat}

If congestion occurs ($\hat{B}_t < R_t$), the bitrate is immediately capped by $\hat{B}_t$. Conversely, when bandwidth is sufficient ($\hat{B}_t \ge R_t$), the algorithm adjusts based on $w_t$. If $C_t > \tau$, $w_t$ becomes negative, causing the bitrate to voluntarily decrease to save bandwidth. When $C_t \approx \tau$, $w_t$ approaches zero, maintaining the bitrate at a stable equilibrium. This mechanism creates a "need-based" transmission pattern, where bandwidth is consumed only when necessary for understanding, leaving a safety margin for bandwidth fluctuations.

\section{Zero-overhead Context-aware Streaming}
\label{sec:roi}


In this section, we propose \textbf{ZeCoStream}. We first describe how to identify response-important regions with zero overhead. Second, we address the challenge of region expiration caused by feedback latency. Third, we present a context-aware bitrate allocation strategy.

\subsection{How to achieve context awareness with zero overhead?}
\label{sec:roi_aware}
To identify response-important regions, a straightforward approach is to deploy lightweight models (e.g., CLIP~\cite{vasu2024mobileclip}) on the client to correlate the conversation context and different video regions. However, as discussed in \S\ref{sec:moti_background}, clients for AI Video Assistants often suffer from limited computation power and battery life (e.g., smart glasses~\cite{ray-ban-meta-2,solo-airgo,inmo,omi-glass}), making it difficult to support local model inference. Although existing solutions propose offloading computation from wearables to paired smartphones, on-device inference on phones still introduces non-negligible computational latency~\cite{vasu2024mobileclip}.

\begin{figure*}[t]
\setlength{\abovecaptionskip}{1.mm}
    \centerline{\includegraphics[width=0.9\linewidth]{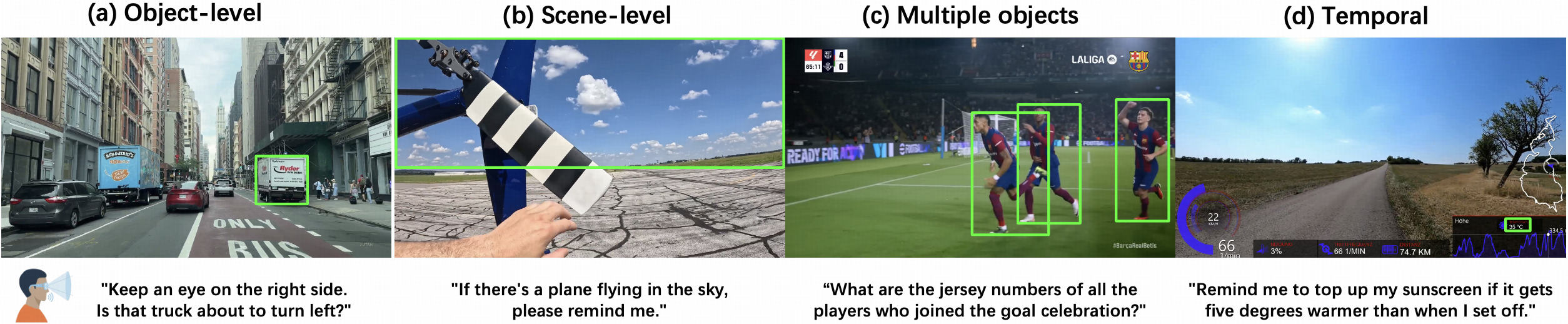}}
    \caption{How to identify response-important regions? Modern MLLMs inherently possess real-time grounding capabilities, enabling them to accurately localize relevant regions (green box) based on the conversational context. Beyond single objects, this capability extends to multiple objects, scene-level contexts, and temporal contexts.}
    \vspace{-0mm}
    \label{fig:grounding}
\end{figure*}

\begin{figure*}
\setlength{\abovecaptionskip}{1.mm}
    \centerline{\includegraphics[width=0.9\linewidth]{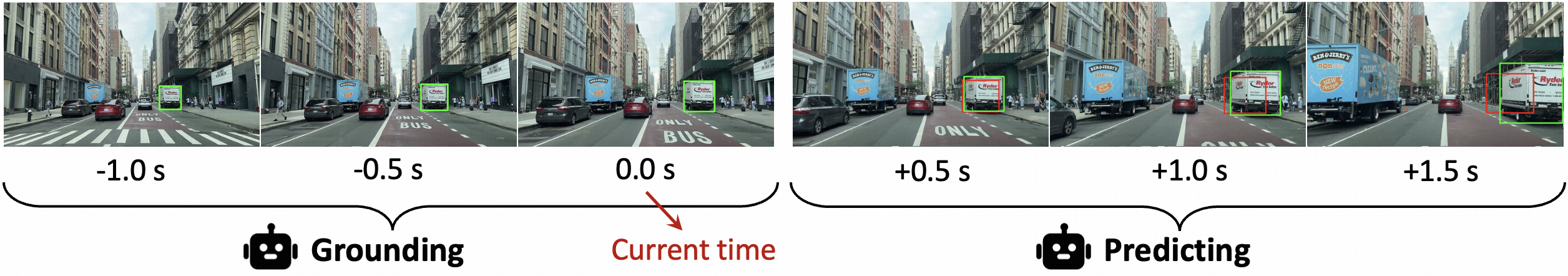}}
    \caption{To offset feedback latency, we instruct the MLLM not only to ground the current important regions (green box) but also to predict their future trajectories (red box) over a short horizon. Details can be found in \S\ref{sec:roi_tracking}.}
    \vspace{-0mm}
    \label{fig:tracking}
\end{figure*}

How can we achieve context awareness with zero overhead? We observe that most state-of-the-art MLLMs inherently possess real-time grounding capabilities~\cite{doubao,glm4.6vflash,glm4.6v,gemini-live,team2024gemini}. As illustrated in Figure~\ref{fig:grounding}, the MLLM can accurately localize relevant regions based on the input prompt. For instance, when the conversation context is ``Keep an eye on the right side. Is that truck about to turn left?'', the truck on the right is precisely localized. Beyond single objects, this capability extends to multiple objects, scene-level contexts, and temporal contexts. Therefore, like ReCapABR, we leverage the server-side MLLM to feed back the bounding boxes of response-important regions to the client in real time.

\subsection{Handling Feedback Latency via Grounding-then-Prediction}
\label{sec:roi_tracking}
However, the feedback regions cannot be directly used for video encoding. This is because these regions are inherently outdated, corresponding to frames captured at least one round-trip time plus the MLLM inference time ago. While this latency is negligible for static objects (e.g., a parked car), it becomes critical for dynamic objects. For instance, a moving vehicle may have shifted out of the outdated bounding box, rendering the feedback useless for the current frame.

To address this, we instruct the MLLM not only to ground the current important regions but also to predict their future trajectories over a short horizon. As shown in Figure~\ref{fig:tracking}, based solely on the past 1.5 seconds of video frames, the MLLM can accurately predict the region's bounding box for the next 1.5 seconds. Thus, a sequence of predicted bounding boxes for multiple future timestamps is sent back to the client. The client then matches the current timestamp with the corresponding prediction to guide video encoding. A key question remains: how long of the latency must the prediction cover? We conducted measurements on a commercial AI Video Assistant (Doubao APP~\cite{doubao}) and found that the total latency from frame capture to receiving the prediction results ranges from 1.20 seconds to 1.52 seconds. Since Figure~\ref{fig:tracking} demonstrates that the MLLM maintains high prediction accuracy within a 1.5-second window, our approach can effectively compensate for the feedback latency.

\subsection{Context-aware bitrate allocation}
\label{sec:roi_control}
With the predicted bounding boxes fed back from the MLLM, the client allocates the limited bitrate to different video regions based on their importance to the response. We divide each video frame ($W \times H$) into a grid of non-overlapping patches (e.g., $64 \times 64$ pixels). Let $\mathcal{K}t$ denote the set of predicted bounding boxes for frame $t$. We calculate the Contextual Importance score $\rho_{i,j} \in [0, 1]$ for the patch at $(i, j)$ as: \begin{alignat}{2}
\rho_{i, j}=\max \left(0,1-\frac{d_{i, j}}{\mu \cdot \sqrt{W^2+H^2}}\right)
\label{equ:importance}
\end{alignat}

where $d_{i,j}$ is the Euclidean distance from the patch center to the nearest boundary of any box in $\mathcal{K}t$ ($d_{i,j}=0$ if inside). We normalize this distance by the frame's diagonal length, scaled by a coefficient $\mu=0.5$. It indicates the importance decays to zero when the distance exceeds half of the diagonal.

We then map this score to the Quantization Parameter (QP) for video encoding. The QP controls the compression level: a larger value reduces bitrate but introduces blurriness, while a smaller value preserves details. Let $Q_{min}$ and $Q_{max}$ denote the lower and upper bounds of the QP range (e.g., $Q_{min}=20, Q_{max}=51$). We adopt a non-linear mapping: \begin{alignat}{2}
Q_{i, j}=Q_{\min }+\left(Q_{\max }-Q_{\min }\right) \cdot\left(1-\rho_{i, j}\right)^2
\label{equ:qp}
\end{alignat}

In this formulation, patches within the important regions ($\rho_{i,j} = 1$) are assigned $Q_{min}$ to ensure maximum fidelity. As the distance increases, the QP rises quadratically towards $Q_{max}$. This forces the encoder to aggressively compress irrelevant regions, ensuring the scarce bitrate budget is concentrated on the content critical for the MLLM's reasoning.

\section{Degraded Video Understanding Benchmark}
\label{sec:bench}

This section proposes \textbf{DeViBench}. We first detail the automatic QA construction, then summarize the benchmark.

\subsection{Automatic QA construction pipeline}
\label{sec:bench_pipeline}
As described in \S\ref{sec:moti_insight}, we need to construct QA samples that are sensitive to video quality. 
For this, the most straightforward way is to hire volunteers to ask tricky questions about degraded videos. However, this is too expensive and inefficient, hindering the scale-up of the dataset. \textbf{So we ask: Can QA samples be constructed automatically and cheaply?} 

Rethinking the background of AI Video Assistant, MLLMs are already capable of understanding videos and giving responses. So we leverage MLLMs to replace human volunteers and develop an automatic QA sample construction pipeline. As illustrated in Figure~\ref{fig:sample_generation}, this pipeline consists of 5 steps:

\begin{figure}
\setlength{\abovecaptionskip}{1.mm}
    \centerline{\includegraphics[width=0.86\linewidth]{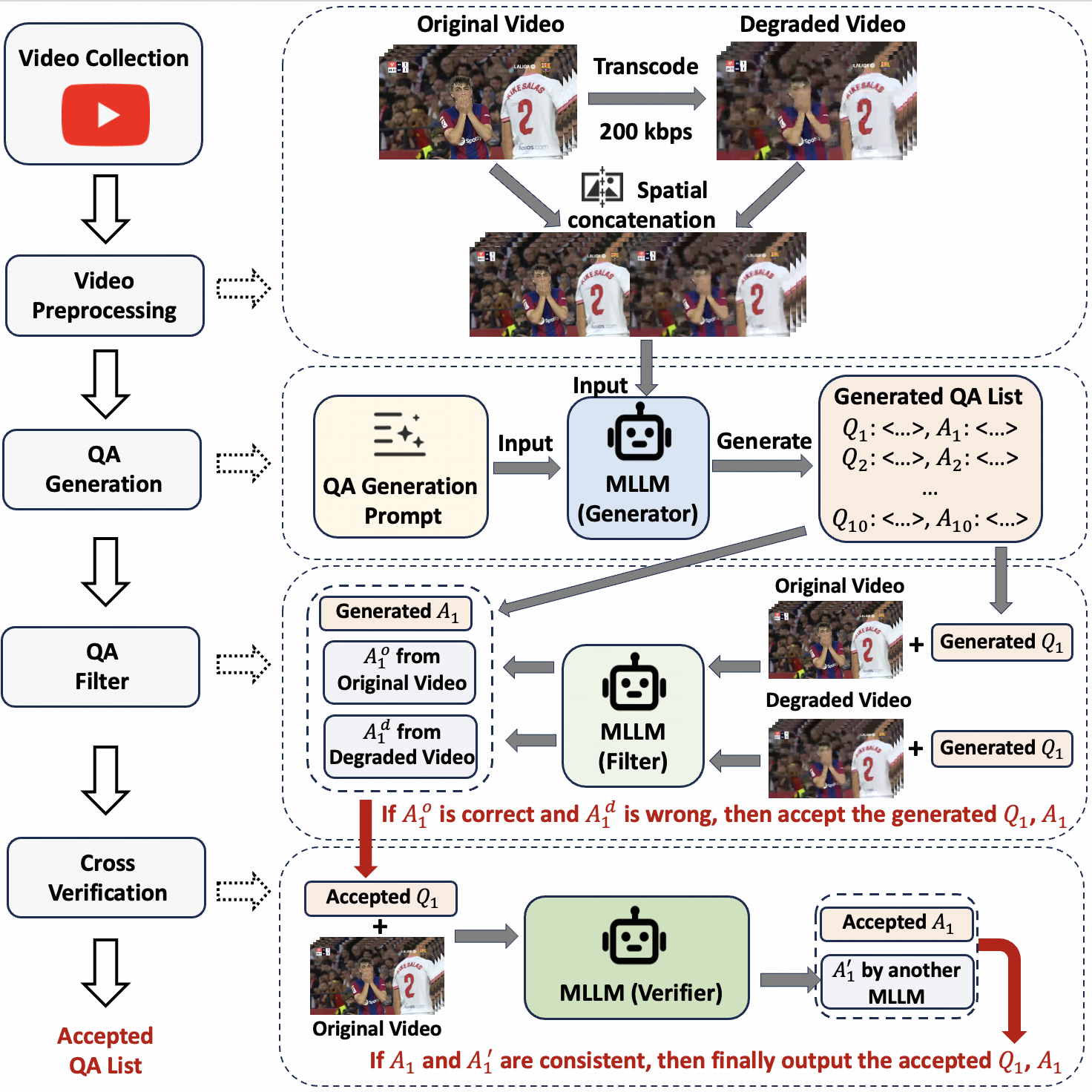}}
    \caption{\wu{DeViBench's pipeline for automatic QA sample construction. Details can be found in \S\ref{sec:bench_pipeline}.}}
    \label{fig:sample_generation}
\end{figure}


\noindent \textbf{Video Collection.} We first collect videos to ask questions. To align with the domain and scale of existing MLLM benchmarks~\cite{lin2024streamingbench}, we directly use their videos (discarding QA).

\noindent \textbf{Video Preprocessing.} To allow MLLMs to understand the quality degradation caused by low bitrates, we transcode the original videos to low-bitrate versions (200 Kbps) using x265 from ffmpeg version N-118035-gc1e3d55f99. The low-bitrate video and the original video are horizontally concatenated into one video. Then this concatenated video will be input to the MLLM for understanding and QA generation. 

\noindent \textbf{QA Generation.} To enable MLLMs to generate QA samples based on the concatenated video, we carefully designed a prompt with guidance from persona, context, core task, execution steps, constraints, and output format, as detailed in the Appendix (Figure~\ref{fig:prompt_generation}). This prompt ensures that MLLMs can recognize quality differences and generate quality-sensitive QA samples. Initially, we experimented with generating multiple-choice questions, but observed that they are inherently less challenging. On one hand, the options embedded in the prompt provide implicit hints to the MLLMs. On the other hand, even when the video quality is severely degraded, MLLMs can still guess from the four candidate options, guaranteeing a baseline accuracy of at least 25\%. Consequently, we shifted to generating free-response questions.
Qwen3-VL-plus~\cite{qwen3vl} is adopted as the generator.


\noindent \textbf{QA Filtering.} The generated QA pairs will be filtered. We separately input the original video and the low bitrate video into the MLLM and use the generated QA pairs for questioning. If the answer from the original video is correct and the answer from the low bitrate video is wrong, we accept this QA pair. Note that since the questions are free-response, exact string matching is inapplicable for assessing correctness. Instead, we employ a separate MLLM as a judge to evaluate the semantic consistency between the response and the answer. In practice, Qwen2.5~Omni~\cite{xu2025qwen2} is used as the filter, while Qwen-Flash~\cite{qwenflash} acts as the judge. In this step, \textbf{25.25\%} of the QA pairs can be accepted.

\noindent \textbf{Cross Verification.} Since the answer generated by generator may also be incorrect, this cannot be filtered out through the above testing. Hence, we utilize another MLLM for cross-verification. We feed the above accepted question into another MLLM, and if the new answer is consistent with the above accepted answer, we finally approve this QA pair. In our experiments, GLM-4.6V~\cite{glm4.6v} is adopted as the verifier and \textbf{89.37\%} of the accepted QA pairs can pass cross-verification. Considering all the above validations together, finally \textbf{22.57\%} of the generated QA pairs are valid.

\subsection{Benchmark Summary}
\label{sec:bench_summary}
Finally, we produce 1,968 QA samples, with details summarized in Table~\ref{tab:benchmark_summary}, including QA types, total duration, total money spent, and total time cost. We also analyze the distribution of different QA types, as shown in Figure~\ref{fig:QA_types}. In terms of categories, there are text-rich understanding (81.86\%), attribute perception (11.53\%), object perception (3.56\%), counting (2.54\%), action perception (0.36\%) and spatial understanding (0.15\%). In terms of temporal dependency, 8.28\% of the questions necessitate inter-frame information (multiple frames) for answering, whereas 91.72\% are answerable with intra-frame information (single frame). The imbalanced data distribution is justifiable, as different questions exhibit varying sensitivities to video degradation. This variation is influenced not only by the question type but also by the dependency on temporal or spatial information.

To confirm whether these MLLM-generated QA samples are usable, we spot-check 100 QA samples for manual answering. Among them, 96\% of the generated questions are answerable by humans, and 93\% of the generated answers are correct. This also proves the effectiveness of our automatic QA construction pipeline. The 100 spot-checked QA samples are designated as the validation set, while the remaining samples constitute the test set. The validation set is used for the hyperparameter tuning of ReCapABR (e.g., $\tau, \gamma$) and ZeCoStream (e.g., $\mu$), as well as for MLLM feedback in-context learning. The test set is used for evaluation (\S\ref{sec:eval_setup}).

\begin{table}[t]
\begin{center}
\setlength{\abovecaptionskip}{2.mm}
\caption{\jiangkai{Benchmark summary}}
\begin{tabular}{rl}
\hline
{\bf \jiangkai{Number of QA samples}}       & \jiangkai{1,968}           \\ \hline
{\bf \jiangkai{QA sample types}}       & \jiangkai{6*2}       \\ \hline
{\bf \jiangkai{Total duration (s)}}    & \jiangkai{88,680}         \\ \hline              
{\bf \jiangkai{Total money spent (\$)}}            & \jiangkai{48.06}         \\ \hline
{\bf \jiangkai{Total time cost (s)}}            & \jiangkai{102,704}     \\ \hline
\end{tabular}
\label{tab:benchmark_summary}
\end{center}
\vspace{-2mm}
\end{table}

\begin{figure}
\setlength{\abovecaptionskip}{2.mm}
    \centerline{\includegraphics[width=0.8\linewidth]{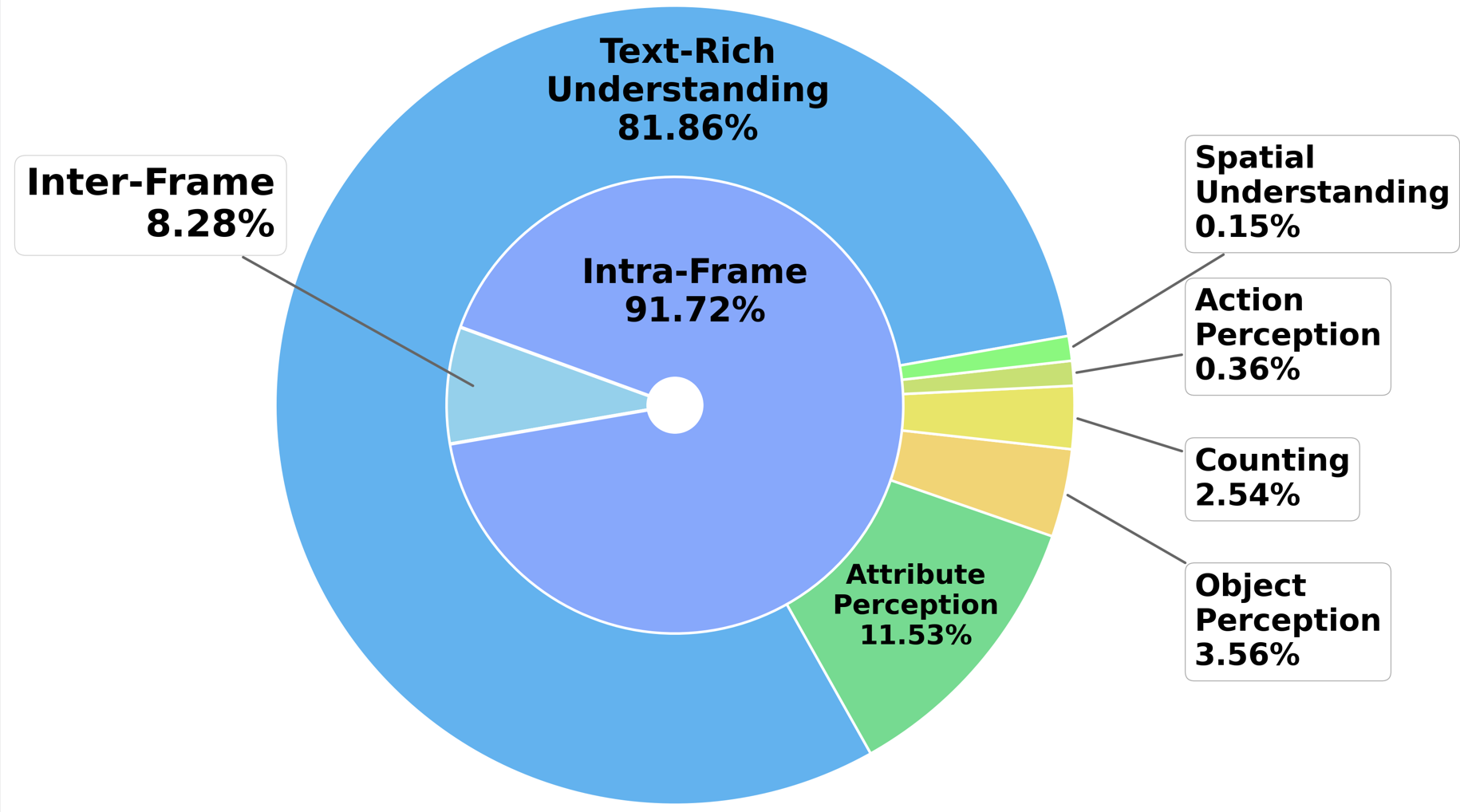}}
    \vspace{1mm}
    \caption{\wu{Distribution of our generated QA samples. Outer ring: QA categories. Inner ring: Whether the question requires inter-frame information to answer.}}
    \vspace{-2mm}
    \label{fig:QA_types}
\end{figure}
\section{Evaluation}
\label{sec:eval}
\subsection{Experimental Setup}
\label{sec:eval_setup}
We built a prototype of Artic in C++ and Python. Here, we describe implementation details and experimental setups.

\noindent\textbf{Baselines.} 
Artic is primarily compared against the traditional human-centric WebRTC, which employs standard video codecs (e.g., H.265~\cite{265}) and conventional CC algorithms (e.g., GCC~\cite{carlucci2016analysis}, BBR~\cite{cardwell2017bbr}). We also establish several strawman baselines to isolate component contributions, including WebRTC+ZeCoStream, WebRTC+ReCapABR.

\noindent\textbf{Metrics.} 
The QoE objectives for AI Video Assistants consist of response accuracy and latency (Table~\ref{tab:qoe_comparison}, \S\ref{sec:moti_diff}). For accuracy, we calculate the ratio of correctly answered samples to the total number of samples on DeViBench (\S\ref{sec:bench}). Note that the validation samples used for hyperparameter tuning are excluded from the calculation. Latency denotes the duration from the client-side encoder to the MLLM-side decoder, termed as frame latency. Frame latency directly contributes to the MLLM's end-to-end response latency. We do not use the full end-to-end latency since it is dominated by MLLM inference time, which is orthogonal to the RTC design.

\noindent\textbf{Test Videos.} 
We conduct evaluations using videos from DeViBench, the scale of which is summarized in Table~\ref{tab:benchmark_summary}. To achieve fine-grained QP control, we adopt the H.265 encoder implemented by Kvazaar~\cite{Kvazaar2016} for both our approach (ZeCoStream) and the baselines (standard encoding). We align the bitrate, resolution, and frame rate configurations with \textit{industry settings}~\cite{agoraencoder}. For decoding, both our method and the baselines utilize x265 via \texttt{decord 0.6.0}, maintaining identical decoding parameters to ensure consistency.

\noindent\textbf{Test MLLMs.} We evaluate accuracy using GLM-4.6V-Flash (9B)~\cite{glm4.6vflash}. To avoid \textit{circularity}, this model is distinct from the filter (Qwen2.5-Omni~\cite{xu2025qwen2}) used to construct DeViBench (\S\ref{sec:bench}). Both the code and the model are frozen prior to testing. We adhere to the default settings (including the same random seed, system prompt, and officially recommended configurations) \textit{without any tuning for the QA test set} (\S\ref{sec:bench_summary}).

\noindent\textbf{Network Traces.} 
To match the traffic pattern of AI Video Assistants, we use a dataset of 5G uplink bandwidth traces~\cite{ghoshal:memu2022}, along with our collected traces (\S\ref{sec:moti_measure}). To reproduce the bandwidth fluctuations caused by user mobility and behavior, we select traces from driving and walking scenarios, filtering segments with significant bandwidth fluctuation. These traces are looped to match the duration of the test videos.

\noindent\textbf{Network Simulation.} 
Building on Razor~\cite{razor}, we developed a WebRTC-based unidirectional video transmission system that supports both GCC and BBR. We employed Mahimahi~\cite{mahimahi} to dynamically limit the sender's uplink bandwidth by replaying the traces described above. We set the queue length to 60 packets and adopted the drop-tail strategy.

\subsection{Latency gains from ReCapABR}
\label{sec:eval_abr}


\newlength{\figH}
\setlength{\figH}{4.2cm}

\begin{figure*} 
  \centering
  \begin{minipage}[t]{0.69\textwidth} 
  \setlength{\abovecaptionskip}{0.mm}
  \centering
  \begin{subfigure}[b]{0.49\linewidth}
  \setlength{\abovecaptionskip}{0.mm}
    \centering
    \includegraphics[width=\linewidth]{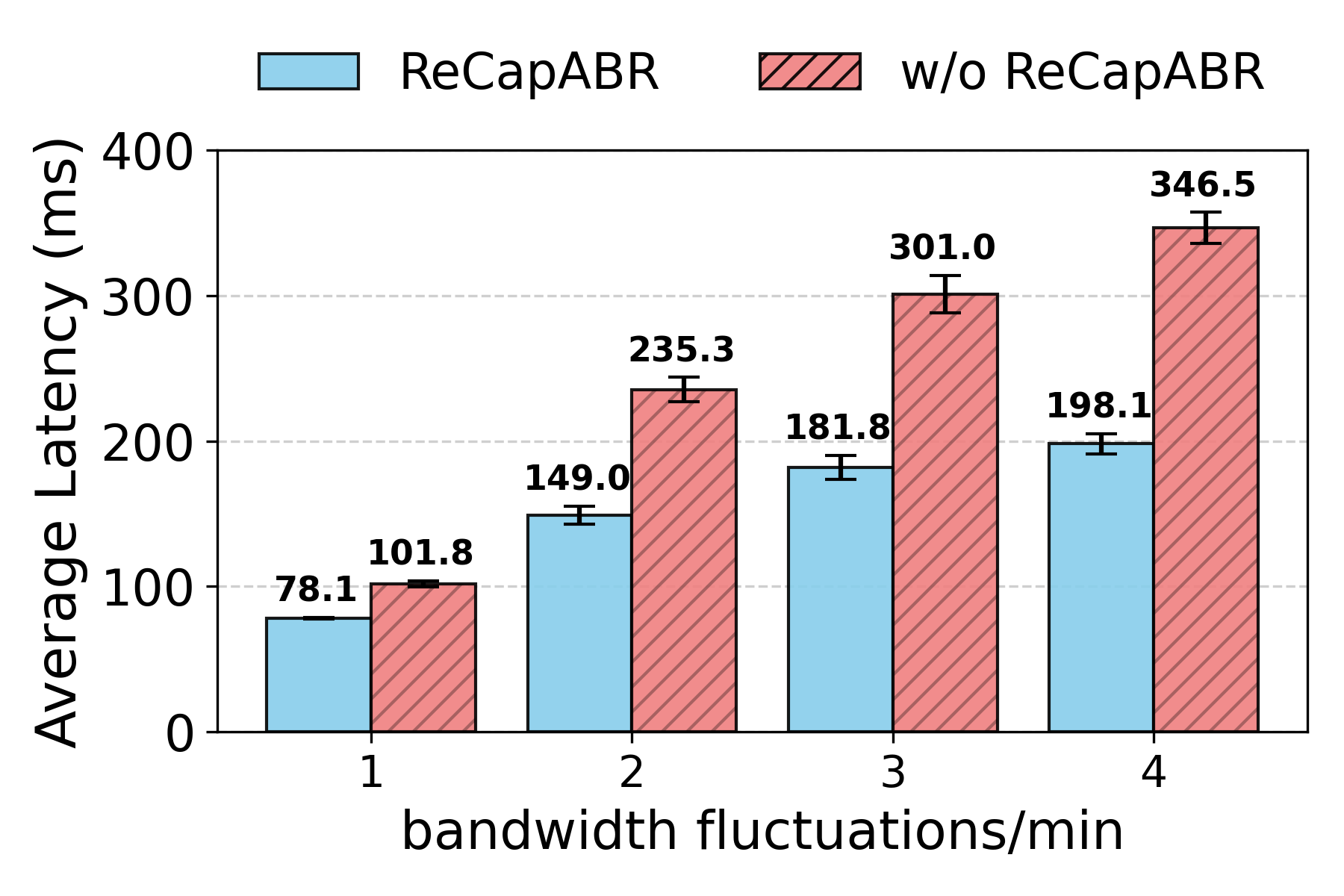}
    \caption{Average latency}
    \label{fig:delay_bar}
  \end{subfigure}
  \begin{subfigure}[b]{0.49\linewidth} 
  \setlength{\abovecaptionskip}{0.mm}
    \centering
    \includegraphics[width=\linewidth]{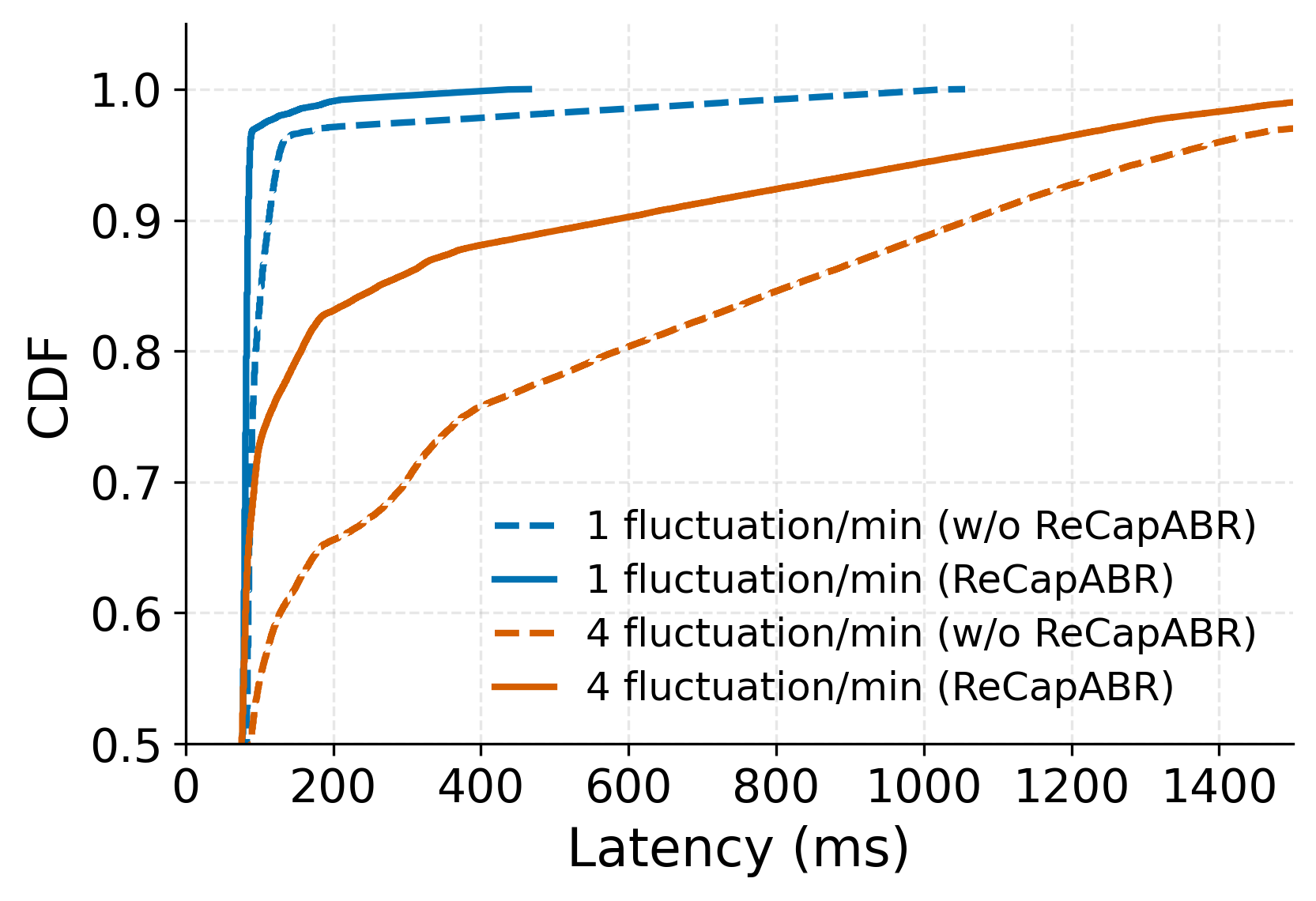}
    \caption{Latency Distribution}
    \label{fig:delay_cdf}
  \end{subfigure}
  \caption{ReCapABR significantly reduces latency under bandwidth fluctuations, especially as fluctuations become more frequent. Details in \S\ref{sec:eval_abr}.}
  \label{fig:delay}
\end{minipage}
  \hfill 
  \begin{minipage}[t]{0.3\textwidth} 
  \setlength{\abovecaptionskip}{0.mm}
    \centering
    \includegraphics[width=\linewidth,height=\figH]{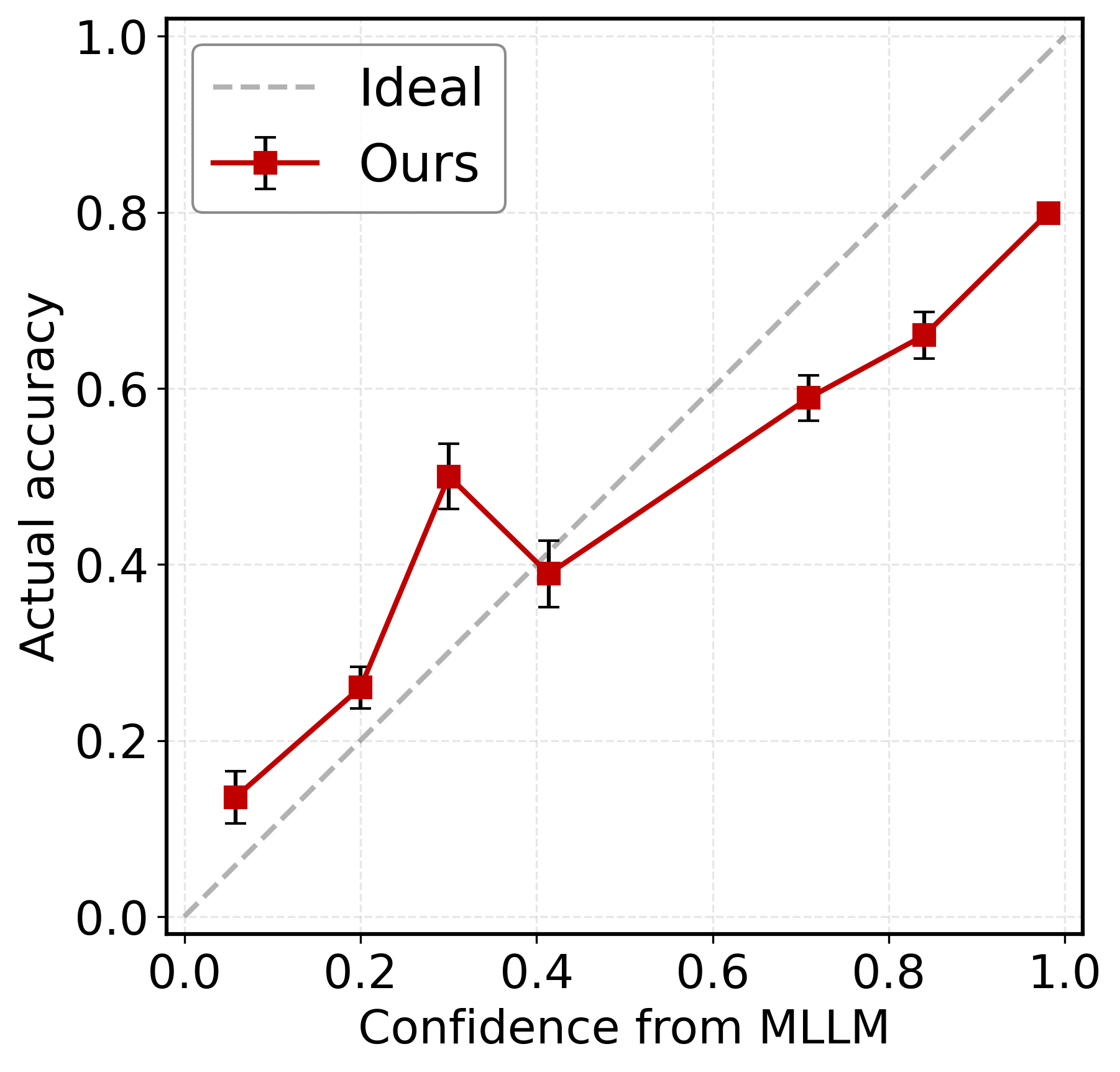}
    \caption{Robustness of ReCapABR to Confidence Errors.}
    \label{fig:robustness}
  \end{minipage}
\end{figure*}

This section demonstrates the latency gains achieved by ReCapABR. We compare ReCapABR against WebRTC (GCC) under various bandwidth fluctuation frequencies. We randomly switch the bandwidth among several industry-standard levels~\cite{agoraencoder} and define fluctuation frequency as the number of switches (fluctuations) per minute. Figures~\ref{fig:delay_bar} and \ref{fig:delay_cdf} show the average latency and the CDF, respectively.

\noindent\textbf{ReCapABR effectively reduces latency under bandwidth fluctuations.} First, ReCapABR consistently reduces average latency across various bandwidth fluctuation frequencies. Second, the more frequent the bandwidth fluctuations, the greater the latency reduction. Figure~\ref{fig:delay_bar} shows that with 1 bandwidth fluctuation per minute, ReCapABR reduces latency by 23.7 ms compared to the baseline. At 4 fluctuations per minute, the reduction expands to 148.4 ms. This is because under intense fluctuations, the baseline, which follows the CC, is more prone to over-transmission. In contrast, by capping the bandwidth estimated by the CC, ReCapABR reserves more headroom to absorb sudden bandwidth drops. Third, ReCapABR significantly increases the proportion of low-latency frames. For example, as illustrated in Figure~\ref{fig:delay_cdf}, at a frequency of 4 fluctuations per minute, only 65.5\% of frames in the baseline have a latency below 200 ms, whereas ReCapABR achieves 82.5\%. This improvement is also attributed to the headroom mechanism of ReCapABR.

\noindent \textbf{Robustness to confidence errors.} ReCapABR relies on the confidence scores from MLLM feedback. Ideally, these scores indicate whether the model can provide correct responses at the current bitrate. Figure~\ref{fig:robustness} illustrates the relationship between confidence scores and actual accuracy. The results show that the confidence scores generally align with the accuracy. Thus, using a conservative capping strategy, ReCapABR remains robust to confidence errors.

\subsection{Accuracy gains from ZeCoStream}
\label{sec:eval_roi}

\begin{figure}
\setlength{\abovecaptionskip}{1.mm}
    \centerline{\includegraphics[width=0.7\linewidth]{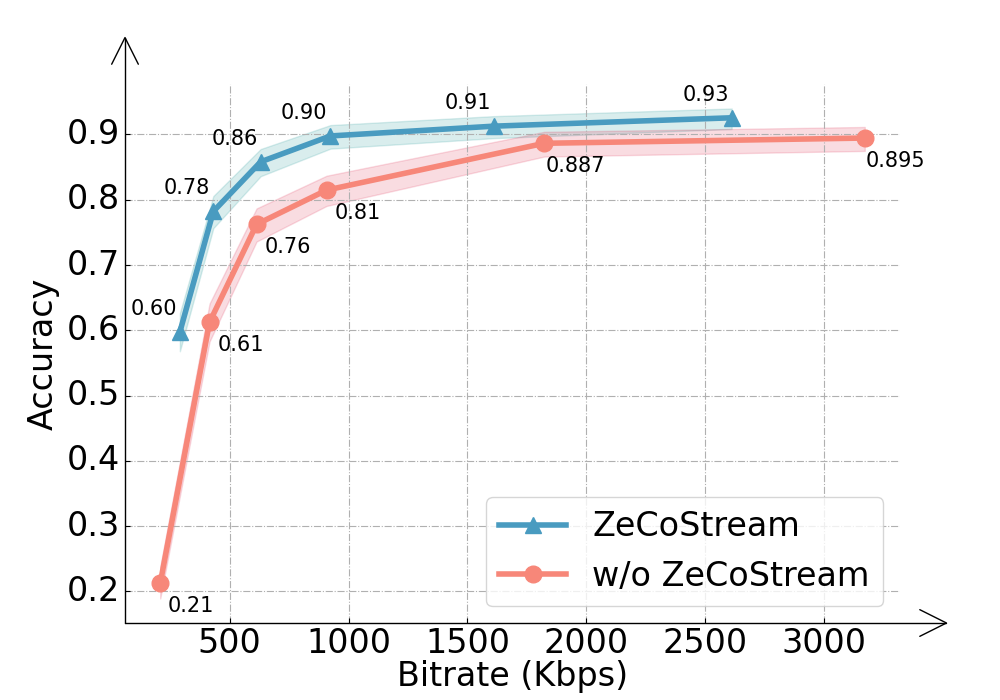}}
    \caption{\jiangkai{ZeCoStream significantly improves accuracy under bandwidth-constrained conditions and reduces the bandwidth required to achieve high accuracy.}}
    \vspace{-4mm}
    \label{fig:accuracy}
\end{figure}

This section proves the accuracy gains of ZeCoStream. Specifically, we compare the accuracy of ZeCoStream against context-agnostic standard encoding across various bitrates (standard industry settings~\cite{agoraencoder}). The results are presented in Figure~\ref{fig:accuracy}.

\begin{figure}
\setlength{\abovecaptionskip}{1.mm}
    \centerline{\includegraphics[width=0.7\linewidth]{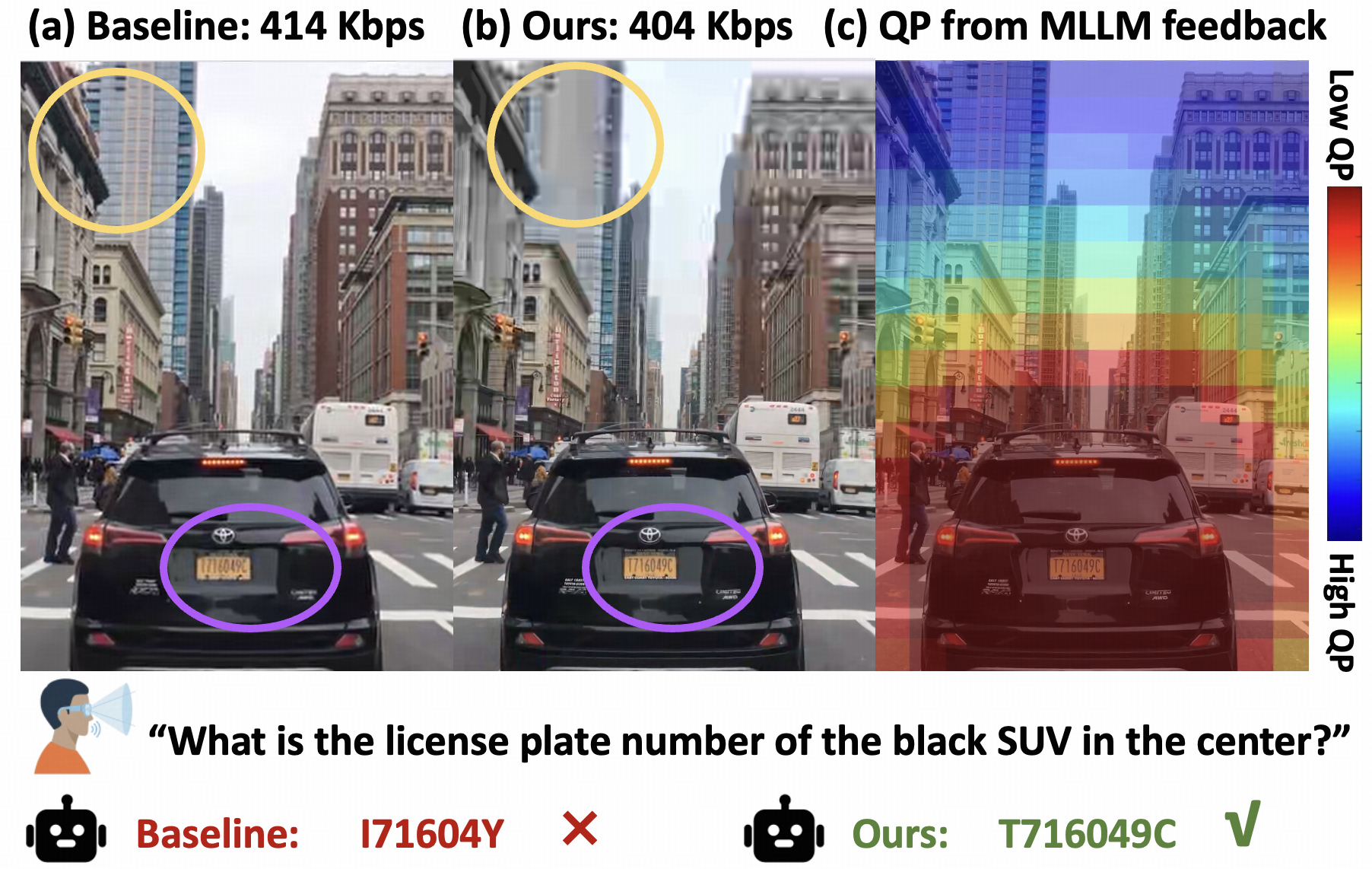}}
    \caption{(a) Encoded with default settings. (b) Encoded with ZeCoStream. (c) The QP from MLLM feedback. The results show that ZeCoStream allocates more bits to important regions (e.g., purple circles) and fewer bits to irrelevant regions (e.g., yellow circles).}
    \vspace{-4mm}
    \label{fig:roi_example}
\end{figure}

\noindent\textbf{ZeCoStream greatly improves accuracy under bandwidth-constrained conditions.} First, in low-bandwidth scenarios, ZeCoStream effectively preserves accuracy. For example, at a low bitrate of 290 Kbps, the baseline accuracy is only 0.39, whereas ZeCoStream maintains it at 0.60. Second, ZeCoStream substantially reduces the bandwidth for high accuracy. For instance, to attain an accuracy of 0.9, the baseline requires a bitrate exceeding 3171.37 Kbps, while ZeCoStream achieves this with only 907.60 Kbps. These gains are attributed to ZeCoStream's ability to allocate limited bitrates to the regions most critical for the response, ensuring that every bit contributes directly to accuracy. Figure~\ref{fig:roi_example} visualizes two sampled frames fed into the MLLM. Even with similar bitrates (414 Kbps vs. 404 Kbps), ZeCoStream preserves higher fidelity in response-important regions (e.g., purple circles) while allowing more blur in response-irrelevant regions (e.g., yellow circles), thus improving MLLM accuracy.

\begin{figure*} 
  \centering
  \begin{minipage}[t]{0.64\textwidth} 
  \setlength{\abovecaptionskip}{0.mm}
  \centering
  \begin{subfigure}[b]{0.49\linewidth}
  \setlength{\abovecaptionskip}{0.mm}
    \centering
    \includegraphics[width=\linewidth]{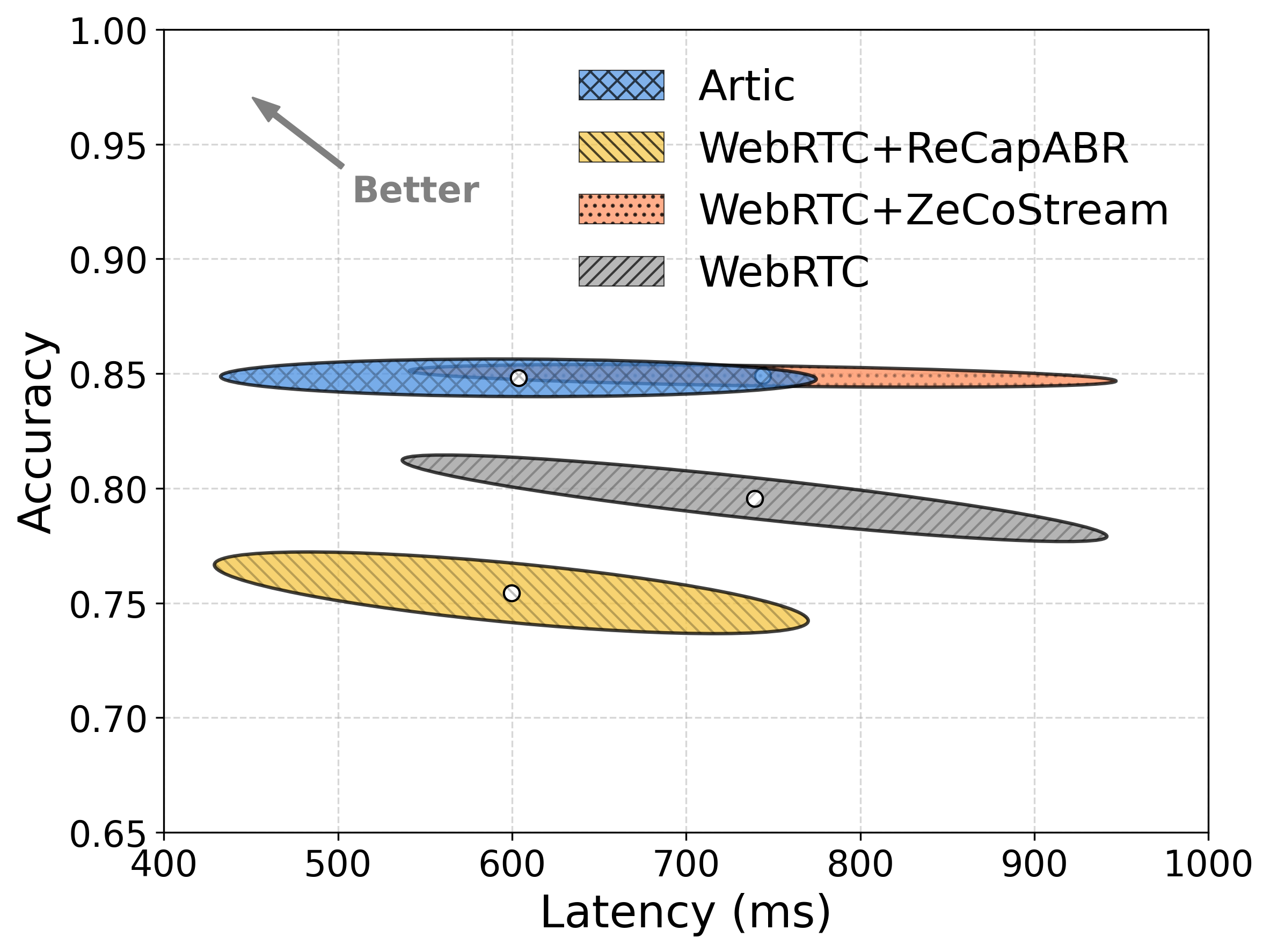}
    \caption{upon BBR}
    \label{fig:qoe_bbr}
  \end{subfigure}
  \begin{subfigure}[b]{0.49\linewidth} 
  \setlength{\abovecaptionskip}{0.mm}
    \centering
    \includegraphics[width=\linewidth]{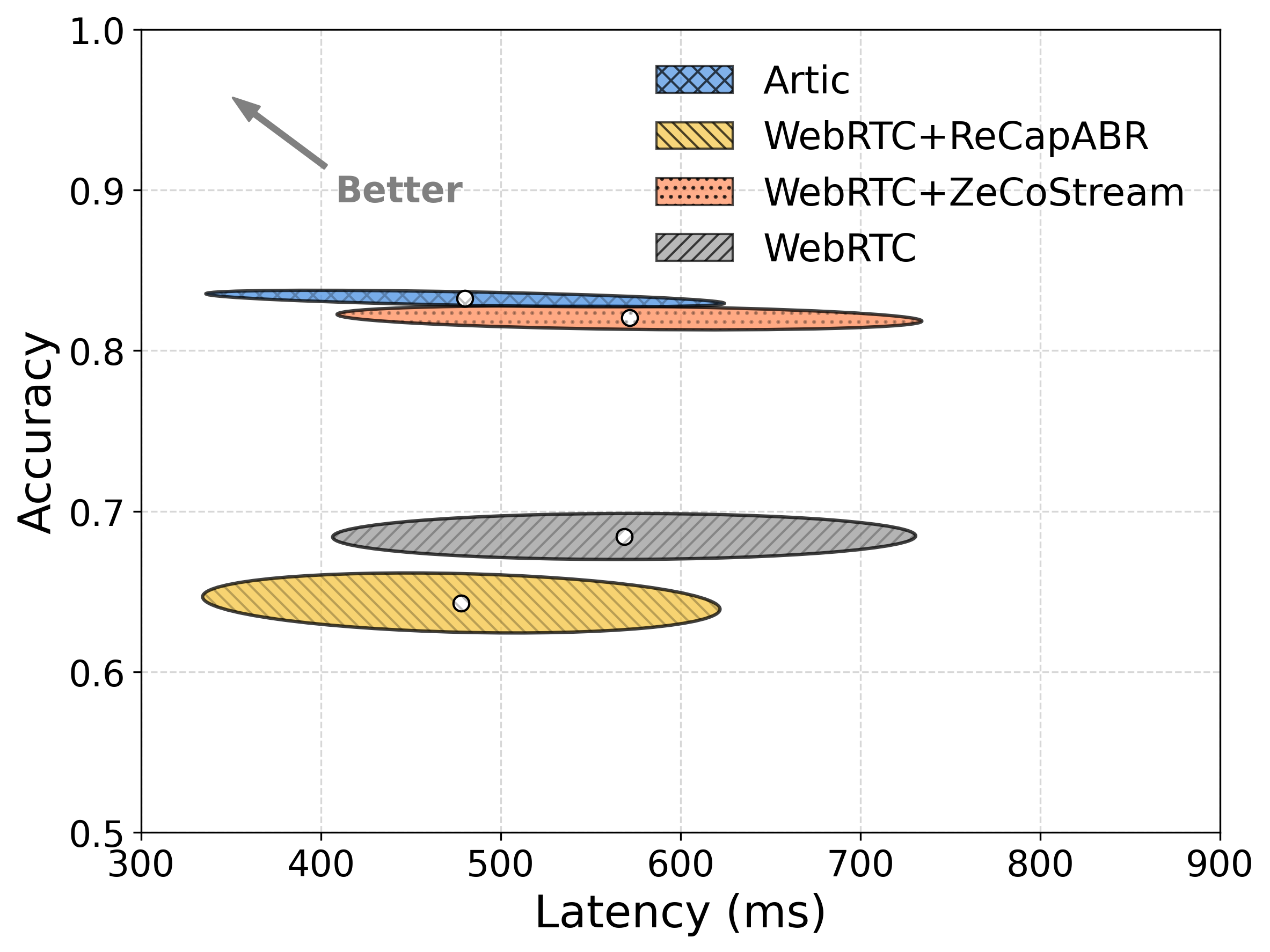}
    \caption{upon GCC}
    \label{fig:qoe_gcc}
  \end{subfigure}
  \caption{The overall performance gains
under real-world network traces. Artic improves accuracy by 15.12\% and reduce latency by 135.31 ms.}
  \label{fig:qoe}
\end{minipage}
  \hfill 
  \begin{minipage}[t]{0.35\textwidth} 
  \setlength{\abovecaptionskip}{0.mm}
    \centering
    \includegraphics[width=\linewidth]{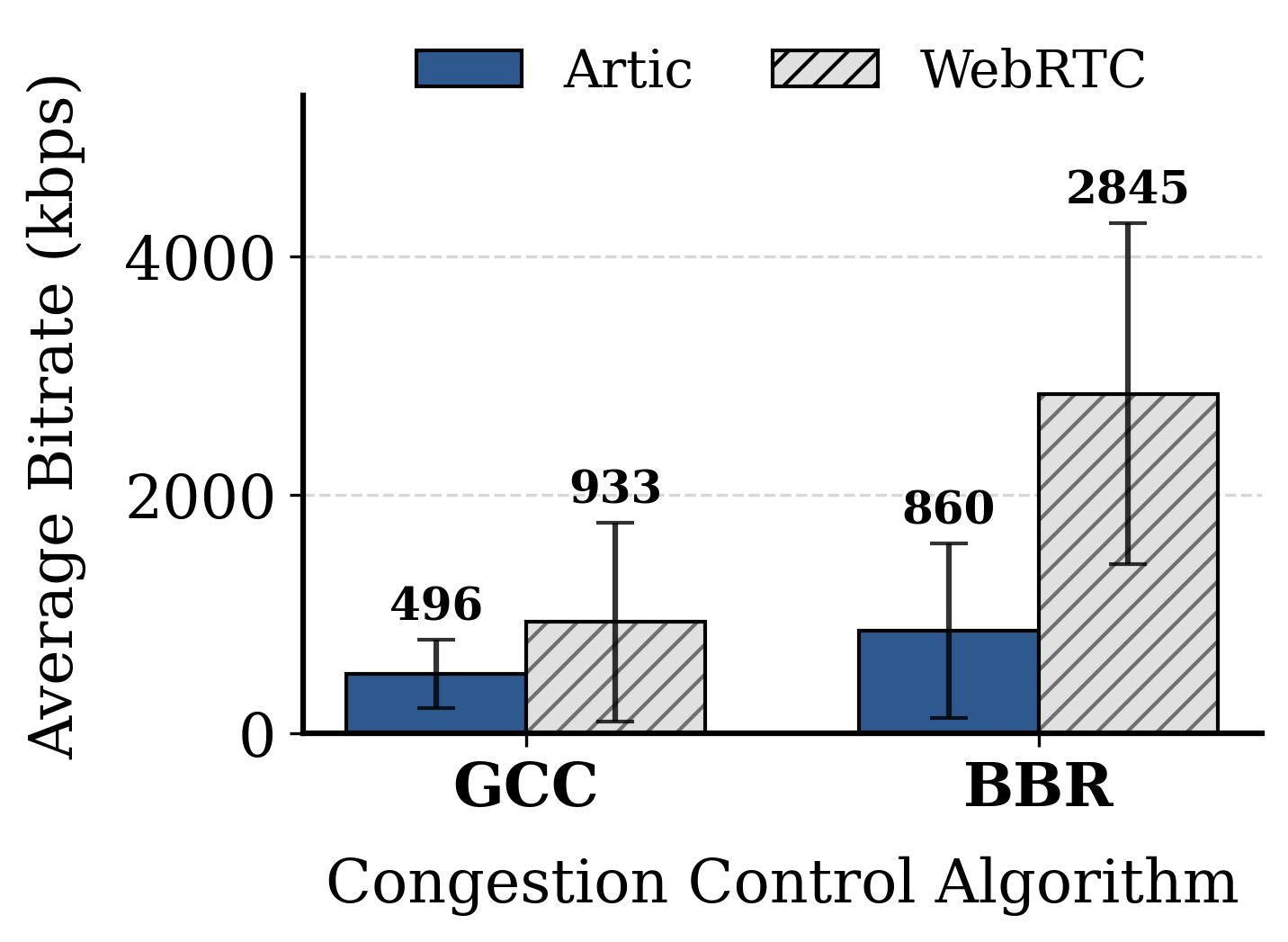}
    \caption{Artic effectively reduces bandwidth overhead by 69.77\%.}
    \label{fig:bitrate_overhead}
  \end{minipage}
\end{figure*}

\subsection{Trace-driven simulation}
\label{sec:eval_overall}

This section demonstrates the end-to-end performance gains under real-world network traces. We compare Artic against three baselines, testing with both GCC and BBR as the underlying CC algorithms. The results are shown in Figure~\ref{fig:qoe}.


\noindent\textbf{Overall performance gains.} First, Artic improves both accuracy and latency. For instance, with BBR (Figure~\ref{fig:qoe_bbr}), Artic boosts the average accuracy from 79.62\% to 84.80\% compared to WebRTC, while reducing the average latency by 135.31 ms. Second, ZeCoStream effectively compensates for the accuracy drop introduced by ReCapABR. For example, with GCC, although WebRTC+ReCapABR reduces latency from 568.13 ms to 477.49 ms compared to WebRTC, it incurs a 4.27\% loss in accuracy (Figure~\ref{fig:qoe_gcc}). Notably, by integrating ZeCoStream, Artic delivers a 19.39\% accuracy gain, surpassing WebRTC by 15.12\%.

\noindent\textbf{Generalization.} 
First, Artic achieves gains in both latency and accuracy across GCC and BBR. Second, we observe that different underlying CC algorithms influence the trade-off between latency and accuracy in Artic. For instance, Artic achieves a 9.94\% higher accuracy gain with GCC compared to BBR, but a 46.82 ms lower latency gain. This is because GCC tends to be conservative and underestimates bandwidth under fluctuations. At low bandwidth estimates, ReCapABR is more likely to CC-based adaptation, resulting in reduced latency gains. Conversely, at lower bitrates, ZeCoStream allocates the limited bitrate to response-important regions, making the accuracy gains more substantial.

\subsection{Overhead}



\begin{figure}
\setlength{\abovecaptionskip}{1.mm}
    \centering
    \includegraphics[width=0.9\linewidth]{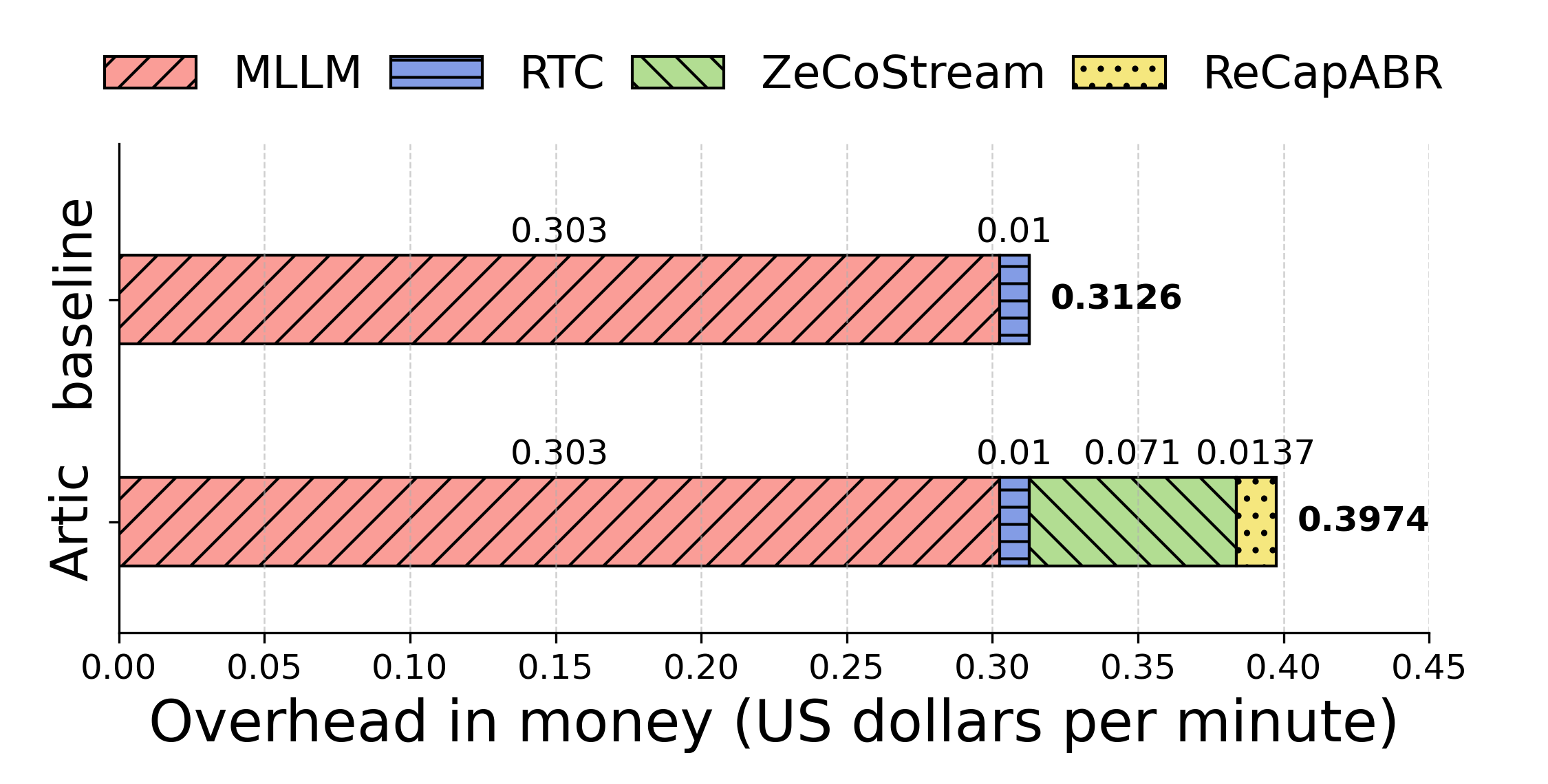}
    \caption{Artic increases monetary cost by 27.13\%.}
    \label{fig:money_overhead}
    \vspace{-3mm}
\end{figure}

In this section, we analyze the overhead of Artic compared to WebRTC for the AI Video Assistant.

\noindent\textbf{Bandwidth Overhead.} 
As discussed in \S\ref{sec:moti_diff}, the bandwidth pressure of the AI Video Assistant lies entirely on the uplink. Artic not only avoids increasing this burden but actually reduces uplink bandwidth overhead via ReCapABR. Figure~\ref{fig:bitrate_overhead} shows Artic reduces bandwidth usage by 46.84\% and 69.77\% (GCC/BBR) vs. WebRTC. This reduction is attributed to the bitrate capping of ReCapABR. On the downlink, feedback overhead is negligible due to minimal data volume (e.g., confidence scores and region coordinates). Furthermore, since there is no video streaming on the downlink, bandwidth is abundant, making the feedback overhead negligible.

\noindent\textbf{Computational Overhead.} 
Artic introduces negligible computational overhead on the client side, as its algorithms primarily rely on feedback from the server-side MLLM (e.g., confidence scores in ReCapABR and important regions in ZeCoStream). Only simple numerical calculations, such as mapping regions to QP and confidence scores to target bitrates, are performed locally; the cost of these operations is negligible. Regarding server-side overhead, since modern MLLMs are typically accessed via APIs and billed based on token usage, we quantify this overhead using monetary cost.

\noindent\textbf{Monetary Cost.} 
How much additional cost does Artic impose on the AI Video Assistant? Based on actual experimental expenditure, we estimated the cost per minute for each component, as shown in Figure~\ref{fig:money_overhead}. For a standard AI Video Assistant, the primary costs arise from the MLLM API (\$0.303/min)~\cite{zhipu_pricing} and the RTC API (\$0.01/min)~\cite{agora_pricing,livekit_pricing}. Artic's ZeCoStream and ReCapABR introduce additional costs of \$0.071/min and \$0.0137/min~\cite{seed_pricing}, respectively. Consequently, Artic increases the total cost from the baseline's \$0.3126/min to \$0.3974/min, a rise of 27.13\%. Given the benefits of a 135.31 ms reduction in latency and a 15.12\% improvement in accuracy (\S\ref{sec:eval_overall}), justifying the cost.

\section{Conclusion}

This paper proposes Artic, an AI-oriented RTC framework for Video Assistants. We design: (1) ReCapABR to absorb bandwidth fluctuations for latency reduction; (2) ZeCoStream to maintain accuracy under low bandwidth; and (3) DeViBench to evaluate how RTC-induced degradation affects MLLM accuracy. Prototype experiments show that compared with existing methods, Artic greatly improves accuracy by 15.12\% and reduces latency by 135.31 ms.

This work does not raise any ethical issues.


\noindent\textbf{Acknowledgements.} This paper is an extended version of our earlier work published in HotNets 2025~\cite{wu2025chat}. Compared to the workshop version, this version includes: (a) A new measurement study (\S\ref{sec:moti_measure}). (b) The entirely new core contribution, ReCapABR (\S\ref{sec:abr}). (c) A redesigned ZeCoStream, which features zero-overhead capabilities (\S\ref{sec:roi}). (d) The refactoring and expansion of DeViBench (\S\ref{sec:bench}). (e) An entirely new evaluation section (\S\ref{sec:eval}).


\bibliographystyle{ACM-Reference-Format}
\bibliography{sigcomm26-reference}

\appendix
\section{Appendix}

\subsection{Prompt used}

\begin{figure} [H]
\centerline{\includegraphics[width=0.95\linewidth]{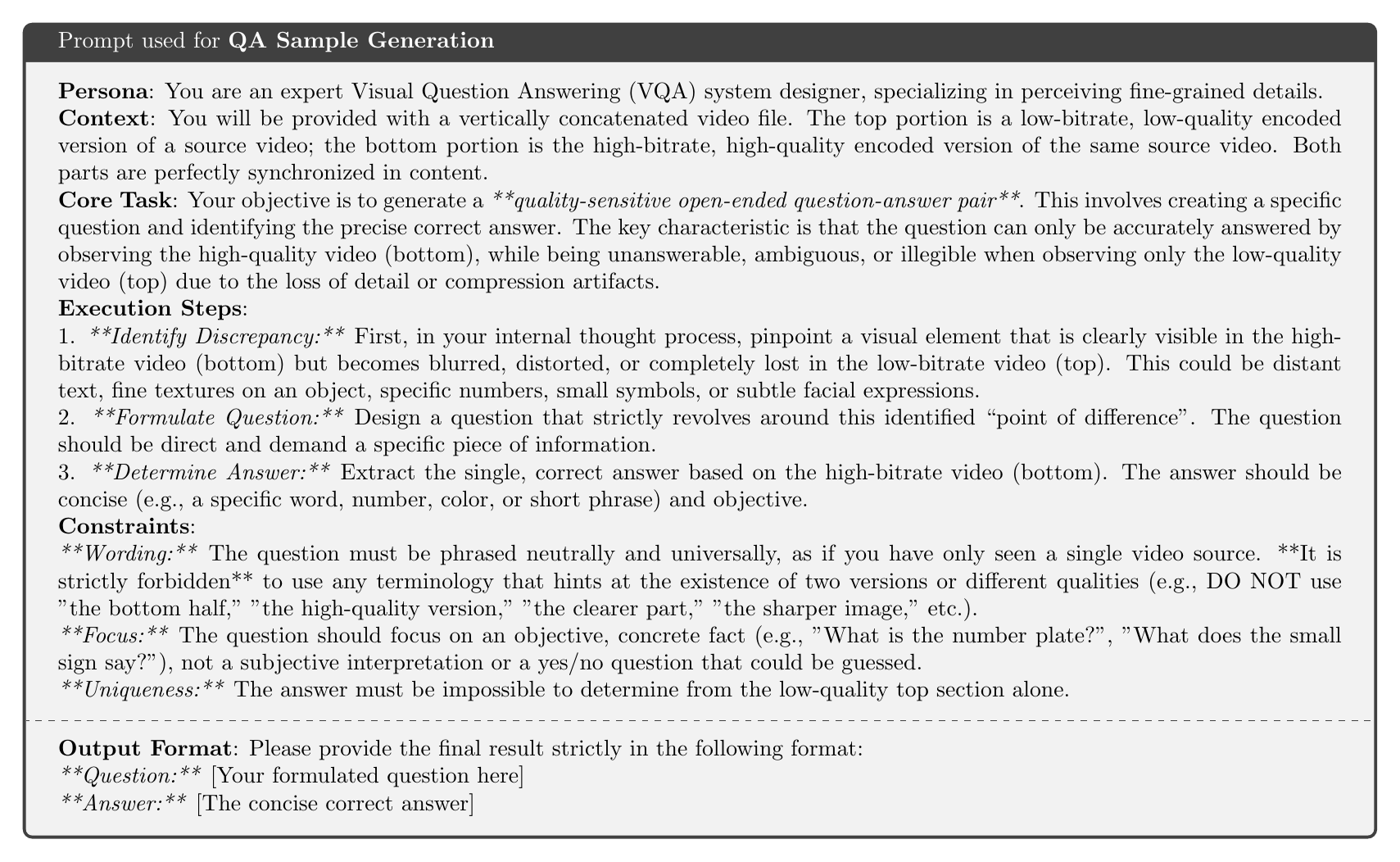}}
    \caption{\wu{Our prompt for QA Sample Generation.}}
    \label{fig:prompt_generation}
\end{figure}

Figure~\ref{fig:prompt_generation} illustrates the prompt used for QA generation.

\end{document}